\documentclass[journal, onecolumn,draft]{IEEEtran}
\IEEEoverridecommandlockouts \makeatletter
\def\ps@headings{%
\def\@oddhead{\mbox{}\scriptsize\rightmark \hfil \thepage}%
\def\@evenhead{\scriptsize\thepage \hfil \leftmark\mbox{}}%
\def\@oddfoot{}%
\def\@evenfoot{}}
\makeatother
\usepackage[dvips]{graphicx}
\usepackage{color}
\usepackage{times}
\usepackage{fancyhdr}
\usepackage{amssymb,amsmath}
\usepackage{lastpage}
\usepackage{epsfig}
\usepackage[lined,boxed,commentsnumbered, ruled]{algorithm2e}
\usepackage[compress]{cite}
\usepackage{cases}
\usepackage{subfigure}
\usepackage{caption}
\usepackage{longtable}
\usepackage{multirow}
\usepackage{threeparttable}
\usepackage{rotating}
\usepackage{bm}
\usepackage{float}
\usepackage{tikz}
\usepackage{array}
\usepackage{cancel}
\usepackage{bbm}

\newcommand{\mc}{\mathcal}

\newtheorem{theorem}{Theorem}
\newtheorem{lemma}{Lemma}
\newtheorem{corollary}{Corollary}
\newtheorem{defn}{Definition}

\newtheorem{example}{Example}
\newcolumntype{C}[1]{>{\centering\let\newline\\\arraybackslash\hspace{0pt}}m{#1}}
\makeatletter
\def\ps@headings{%
\def\@oddhead{\mbox{}\scriptsize\rightmark \hfil \thepage}%
\def\@evenhead{\scriptsize\thepage \hfil \leftmark\mbox{}}%
\def\@oddfoot{}%
\def\@evenfoot{}}
\makeatother

\begin{document}
\title{On Zero-Error Capacity of Graphs with One Edge}

\author{Qi Cao, Qi Chen and Baoming Bai
\thanks{Qi Cao (caoqi@xidian.edu.cn) is with Xidian-Guangzhou Research
  Institute, Xidian University, Guangzhou, China.
Qi Chen (qichen@xidian.edu.cn) is with the School of
  Telecommunications Engineering, Xidian University, Xi'an 710071,
  China. 
Baoming Bai (bmbai@mail.xidian.eu.cn) is with the State Key Laboratory of
  Integrated Service Networks, Xidian University, Xi'an 710071,
  China.}
\thanks{This paper was presented in part at ISIT2022\cite{9834528}.} 
}
\maketitle

\normalsize 


\begin{abstract}
In this paper, we study the zero-error capacity of channels with
memory, which are represented by graphs. We provide a method to construct code for any graph with one edge, thereby determining a lower bound on its zero-error capacity. Moreover, this code can achieve zero-error capacity when the symbols in a vertex with degree one are the same. We further apply our method to the one-edge graphs representing the binary channels with two memories. There are 28 possible graphs, which can be organized into 11 categories based on their symmetries. The code constructed by our method is proved to achieve the zero-error capacity for all these graphs except for the two graphs in Case 11.
\end{abstract}
\begin{IEEEkeywords}
zero-error capacity, graph with one edge, channel with memory
\end{IEEEkeywords}

\section{Introduction}\label{sec:introduction}
Let $\mc{X}$ be a finite set. A channel with
transition matrix $p(z|x), x\in \mc{X}$ is represented by a graph
$G=(\mc{X},E)$, where the vertex set is $\mc{X}$ and the edge set is
$E$ with $uv\in E$ for $u,v\in\mc{X}$ if 
$$\{z:p(z|u)\}\cap \{z:p(z|v)\}=\emptyset.$$
For $G$, $u$ and $v$ is called \emph{distinguishable} if $uv\in E$. Any two sequences $\bm{x}, \bm{y}\in \mc{X}^n$ is also called
distinguishable if there exists a coordinate $i$, the $i$-th symbols
of $\bm{x}$ and $\bm{y}$ are adjacent in $G$ \cite{1956,zero-noisy2022}.

Now for a set $\mc{A}_n\subseteq \mc{X}^n$ of $|\mc{X}|$-ary sequences, if
sequences in $\mc{A}_n$ are pairwise distinguishable, then messages mapping to
$\mc{A}_n$ can be transmitted through $G$ without error. The set $\mc{A}_n$ is called a code of length $n$ for $G$ and
$\frac{1}{n}\log|\mc{A}_n|$ is its \emph{rate}, where the base of the logarithm is 2, which is omitted throughout this paper. Let $\{\mc{A}_n\}$ be a
sequence of the codes for $G$. The \emph{zero-error capacity} of $G$ is defined to be the maximum among all
\begin{equation*}
\lim_{n\to\infty}\frac{1}{n}\log|\mc{A}_n|.
\end{equation*}

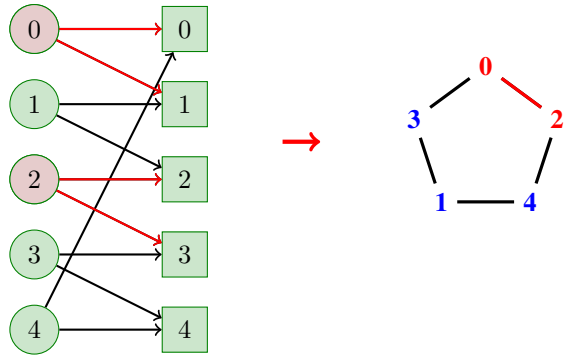
\begin{figure}
\centering
\begin{tikzpicture}
    \node [circle,fill=green!50!black!20, draw=green!50!black,inner sep=3pt,minimum size=6.5mm,] (a) at (0,0) {$0$};
    \node [circle,fill=green!50!black!20, draw=green!50!black,inner sep=3pt,minimum size=6.5mm,] (b) at (0,-1) {$1$};
    \node [circle,fill=green!50!black!20, draw=green!50!black,inner sep=3pt,minimum size=6.5mm,] (c) at (0,-2) {$2$};
    \node [circle,fill=green!50!black!20, draw=green!50!black,inner sep=3pt,minimum size=6.5mm,] (d) at (0,-3) {$3$};
    \node [circle,fill=green!50!black!20, draw=green!50!black,inner sep=3pt,minimum size=6.5mm,] (e) at (0,-4) {$4$};
    \node [rectangle,fill=green!50!black!20, draw=green!50!black,inner sep=3pt,minimum size=6mm,] (A) at (2,0) {$0$};
    \node [rectangle,fill=green!50!black!20, draw=green!50!black,inner sep=3pt,minimum size=6mm,] (B) at (2,-1) {$1$};
    \node [rectangle,fill=green!50!black!20, draw=green!50!black,inner sep=3pt,minimum size=6mm,] (C) at (2,-2) {$2$};
    \node [rectangle,fill=green!50!black!20, draw=green!50!black,inner sep=3pt,minimum size=6mm,] (D) at (2,-3) {$3$};
    \node [rectangle,fill=green!50!black!20, draw=green!50!black,inner sep=3pt,minimum size=6mm,] (E) at (2,-4) {$4$};

    \path [thick,->]
    (a) edge (A)
    (b) edge (B)
    (c) edge (C)
    (d) edge (D)
    (e) edge (E)
    (a) edge (B)
    (b) edge (C)
    (c) edge (D)
    (d) edge (E)
    (e) edge (A);
    
        \node [circle,fill=red!50!black!20, draw=green!50!black,inner sep=3pt,minimum size=6.5mm,] (a) at (0, 0) {$0$};
        \node [circle,fill=red!50!black!20, draw=green!50!black,inner sep=3pt,minimum size=6.5mm,] (c) at (0,-2) {$2$};
        \path [thick,->,red]
        (a) edge (A)
        (c) edge (C)
        (a) edge (B)
        (c) edge (D);
        
    {\draw[->,ultra thick,red]  (3.3,-1.5) ->(3.8,-1.5);}
    {
        \node [] (aa) at (6,   -0.5) {\textcolor[rgb]{1,0,0}{\textbf{0}}};
        \node [] (bb) at (5.05,-1.2) {\textcolor[rgb]{0,0,1}{\textbf{3}}};
        \node [] (cc) at (5.41,-2.3) {\textcolor[rgb]{0,0,1}{\textbf{1}}};
        \node [] (dd) at (6.59,-2.3) {\textcolor[rgb]{0,0,1}{\textbf{4}}};
        \node [] (ee) at (6.95,-1.2) {\textcolor[rgb]{1,0,0}{\textbf{2}}};
        \path [-,very  thick](aa) edge (bb) edge (ee);
        \path [-,very  thick](cc) edge (bb) edge (dd);
        \path [-,very  thick](dd) edge (ee);
        \path [red,-,very  thick](aa) edge (ee);
    }
\end{tikzpicture}
\caption{Typewriter channel.}\label{fig-type}
\end{figure}

The zero-error capacity problem was introduced by Shannon
\cite{1956} in 1956. 
He considered a typewriter channel, as shown in Fig.~\ref{fig-type}, which can be represented by a graph with length $5$, i.e., each $x\in\mc{X}$ with $|\mc{X}|=5$ is distinguishable from another
two elements in $\mc{X}$. He established a lower bound of the capacity, which was proved tight by Lov{\'a}sz~\cite{1979} in
1979. The problem remains open even for the complement of a cycle
graph with length 7.\footnote{The zero-error capacity problem is trivial for the
complement of a cycle graph with even length.}
Due to the difficulties in solving the problem in general, in recent
years, the zero-error capacity of some special graphs were
investigated.
In 2010, Zhao and Permuter \cite{Zhao} introduced a dynamic programming formulation for computing the zero-error feedback capacity of channels with state information.
The zero-error capacity of some special timing channels were
determined by Kova{\v{c}}evi{\'c} and Popovski \cite{Kovacevic} in 2014.
In 2016, Nakano and Wadayama \cite{NAkano} derived a lower bound and an upper bound on the zero-error capacity of Nearest Neighbor Error channels with a multilevel alphabet.

The zero-error capacity of a channel with memory was first studied by
Ahlswede \emph{et al.} \cite{1998} in 1998. A channel with $m$
memories can be represented by $G$ with $V(G)=\mc{X}^{m+1}$.
In \cite{1998}, authors studied a binary channel with one memory
i.e., a channel represented by $G$ with $V(G)=\mc{X}^{2}$, where
$\mc{X}=\{0,1\}$, and any two vertices may or may not be
distinguishable. 
They determined the zero-error capacity when only one pair of the vertices are distinguishable.
Based on their work, Cohen \emph{et al.} \cite{2014b} in 2016 studied channels with 3 pairs of vertices being distinguishable.
All the remaining cases were solved in 2018 by Cao
\emph{et~al}~\cite{Cao2018}.

However, when $m>1$ or $|\mc{X}|>2$, the number of cases will explode dramatically, making it completely impossible to be solved one by one. For example, when $m=1$ and $|\mc{X}|=3$, the number of cases is $2^{36}\approx 68$ billion. 
There arises a pressing demand for a generalized result. This paper considers any graph with $\bm{u}\bm{v}$ the only edge, where $\bm{u},\bm{v}\in \mathcal{X}^{m+1}$, $m\ge 1$ and $\mc{X}$ is a finite set. This graph is denoted by $G(\bm{u},\bm{v})$ or $G(\bm{v},\bm{u})$.
We devise a simple method to establish a code for $G(\bm{u},\bm{v})$, thus obtaining a lower bound on its zero-error capacity.
For any graph $G$ with more than one edge, let $\bm{u}\bm{v}$ be one of its edges. Note that any code for $G(\bm{u},\bm{v})$ is also the code for $G$. Our method is applicable for establishing a code for any non-empty graph, thus obtaining a general lower bound on zero-error capacity.

We apply our method to the binary channels with two memories represented by the graphs with only one edge.
There are 28 possible graphs, which can be classified into 11 categories up to symmetry (see Section~II for more details).
The capacities of the graphs in each category are the same, so only one
of them need to be considered.
Table~\ref{table} summarizes all the solved and unsolved cases. The code constructed by our method achieves the zero-error capacity for all these graphs except for the two graphs in Case 11.

\begin{table}
\renewcommand\arraystretch{1.1}
\caption{Zero-error capacity of binary channels with two memories }\label{table}
\newcommand{\tabincell}[2]{\begin{tabular}{@{}#1@{}}#2\end{tabular}}
\centering
\begin{threeparttable}
    \begin{tabular}{|C{0.7cm}|C{1.8cm}|C{2.2cm}|C{1.5cm}|}			
        \hline
        Case & $G$ & $C(G)$ & Theorems\\
        \hline
        \multirow{4}{*}{1}
        & $G(000,001)$ & \multirow{4}{*}{${-\log \alpha \approx 0.551}$\tnote{1}} & \multirow{4}{*}{Theorem~\ref{thm:1}}\\
        \cline{2-2}
        & $G(000,100)$ & & \\
        \cline{2-2}
        & $G(111,110)$ & & \\
        \cline{2-2}
        & $G(111,011)$ & & \\
        \hline
        \multirow{2}{*}{2}
        & $G(000,010)$ & \multirow{2}{*}{$\frac{1}{2}$}& \multirow{2}{*}{Theorem~\ref{thm:2}}\\
        \cline{2-2}
        & $G(111,101)$ & & \\
        \hline
        \multirow{4}{*}{3}
        & $G(000,011)$ & \multirow{12}{*}{${-\log \beta \approx 0.406}$\tnote{2}}& \multirow{4}{*}{Theorem~\ref{thm:3}}\\
        \cline{2-2}
        & $G(000,110)$ & & \\
        \cline{2-2}
        & $G(111,100)$ & & \\
        \cline{2-2}
        & $G(111,001)$ & & \\
        \cline{1-2}
        \cline{4-4}
        \multirow{4}{*}{4}
        & $G(010,011)$ & & \multirow{4}{*}{Theorem~\ref{thm:4}}\\
        \cline{2-2}
        & $G(010,110)$ & & \\
        \cline{2-2}
        & $G(101,100)$ & & \\
        \cline{2-2}
        & $G(101,001)$ & & \\
        \cline{1-2}
        \cline{4-4}
        \multirow{4}{*}{5}
        & $G(010,001)$ & & \multirow{4}{*}{Theorem~\ref{thm:5}}\\
        \cline{2-2}
        & $G(010,100)$ & & \\
        \cline{2-2}
        & $G(101,110)$ & & \\
        \cline{2-2}
        & $G(101,011)$ & & \\
        \hline
        6& $G(000,111)$ & \multirow{8}{*}{$\frac{1}{3}$}& {Theorem~\ref{thm:6}}\\
        \cline{1-2}
        \cline{4-4}
        7& $G(010,101)$ & & {Theorem~\ref{thm:7}}\\
        \cline{1-2}
        \cline{4-4}
        \multirow{2}{*}{8}
        & $G(100,011)$ & & \multirow{2}{*}{Theorem~\ref{thm:8}}\\
        \cline{2-2}
        & $G(110,001)$ & & \\
        \cline{1-2}
        \cline{4-4}
        \multirow{2}{*}{9}
        & $G(000,101)$ & & \multirow{2}{*}{Theorem~\ref{thm:9}}\\
        \cline{2-2}
        & $G(111,010)$ & & \\
        \cline{1-2}
        \cline{4-4}
        \multirow{2}{*}{10}
        & $G(001,011)$ & & \multirow{2}{*}{Theorem~\ref{thm:10}}\\
        \cline{2-2}
        & $G(110,100)$ & & \\
        \hline
        \cline{4-4}
        \multirow{2}{*}{11}
        & $G(001,100)$
        & \multirow{2}{*}{Not Determined\tnote{3} } & \multirow{2}{*}{Theorem~\ref{thm:11}}\\
        \cline{2-2}
        & $G(110,011)$ & & \\
        \hline
    \end{tabular}
    \begin{tablenotes}
        \item[1] $\alpha$ is the positive root of the equation $x  +x^3=1$.
        \item[2] $\beta $ is the positive root of the equation $x^2+x^3=1$.
        \item[3] $C(G)\in [\frac{\log 14}{11}\approx 0.346, -\log \beta\approx 0.406]$.
    \end{tablenotes}
\end{threeparttable}
\end{table}

\section{Zero-Error Capacity Problem with Memories}\label{sec:model}
For a finite set $\mathcal{X}$ of symbols, the channel with $m-1$ memories can be
represented by a graph $G=(V,E)$,
where $m\ge 2$, $V={{\mathcal{X}}^{m}}$, and
for any $\bm{u},\bm{v}\in V(G)$, $\bm{u}\bm{v}\in E$ if
$$\{\bm{z}:p(\bm{z}|\bm{u})\}\cap
\{\bm{z}:p(\bm{z}|\bm{v})\}=\emptyset, 
$$
and $\bm{u}$ and $\bm{v}$ are called \emph{distinguishable} for $G$.

For $n_1,n_2\in \mathbb{Z}$, $n_1\le n_2$, let $\mathbb{Z}[n_1,n_2]=\{i\in\mathbb{Z}:n_1\le i\le n_2\}$.
For $\bm{x}=(x_0,x_1,\ldots,x_{n-1}), \bm{y}=(y_0,y_1,\ldots,y_{n-1})\in \mathcal{X}^n$,
$n\ge m$,
we say $\bm{x}$ and $\bm{y}$ are \emph{distinguishable} for $G$ if there exists at least one coordinate
$i\in \mathbb{Z}[0,n-m-1]$ with \[\{x_ix_{i+1},\ldots,x_{i+m},y_iy_{i+1},\ldots,y_{i+m}\} \in E(G).\]

\begin{defn}
Let $\mathcal{A}_n$ be a set of length~$n$ sequences and $\{\mathcal{A}_n\}$, $n=m,m+1,\ldots$, be a sequence of such sets indexed by $n$.
The {\em asymptotic rate} of $\{\mathcal{A}_n\}$ is $R(\{\mathcal{A}_n\})\triangleq\underset{n\to \infty }{\mathop{\lim}}\, \frac{1}{n}\log |\mathcal{A}_n|$ if it exists. If the sequences in $\mathcal{A}_n$
are pairwise distinguishable for the graph $G$, then $\mathcal{A}_n$ is called a {\em code} of length $n$ for $G$,
and the sequences in $\mathcal{A}_n$ are called {\em codewords}.

\end{defn}

\begin{defn}
Let $\{\hat{\mathcal{A}}_n\}$ be a sequence of codes for the graph $G$ such that for all $n$, $\hat{\mathcal{A}}_n$ achieves the largest cardinality of a code of length~$n$ for $G$.
The \emph{zero-error capacity} of the graph $G$ is defined as
\[C(G)=\underset{n\to \infty }{\mathop{\lim}}\, \frac{1}{n}\log |\hat{\mathcal{A}}_n|.\]
\end{defn}
Note that the limit above always exists because $\log |\hat{\mathcal{A}}_n|$ is superadditive, i.e.,
$\log |\hat{\mathcal{A}}_{m+n}|\ge \log |\hat{\mathcal{A}}_m|+\log |\hat{\mathcal{A}}_n|,$ $\forall m,n\ge 0.$
Clearly, $0\le C(G)\leq \log |\mathcal{X}|$.
A sequence of codes $\{{\mathcal{A}}_n\}$ is said to be \emph{asymptotically optimal} for the graph $G$ if $R(\{{\mathcal{A}}_n\})=C(G)$.

We apply the method in \cite{Cao2018} to construct a new code based on an existed code.
Let ${\mathcal{A}}_n$ be a set of length~$n$ sequences and $\{{\mathcal{A}}_n\}$ be a sequence of such sets indexed by $n$. For any $n$, by adding an arbitrary
prefix ${\bm{s}}_\mathrm{p}$ of length $p\ge0$ and an arbitrary suffix ${\bm{s}}_\mathrm{s}$
of length $l\ge 0$ to all the sequences in
${\mathcal{A}}_n$, we obtain a new set of sequences of length $t\triangleq n+(p+l)$,
denoted by $\mathcal{A}'_t$. Let $\{\mathcal{A}'_t\}$ be a sequence of such sets indexed by $t$.
\begin{lemma}[Lemma 2 in \cite{Cao2018}]\label{lm1}
$R(\{{\mathcal{A}}_n\})=R(\{\mathcal{A}'_t\})$, i.e., $\underset{n\to \infty }{\mathop{\lim }}\,\frac{1}{n}\log \left| {{\mathcal{A}}_n} \right|=\underset{t\to \infty }{\mathop{\lim }}\,\frac{1}{t}\log \left| {{\mathcal{A}}'_t} \right|$.
\end{lemma}
Obviously, if $\{{\mathcal{A}}_n\}$ is a sequence of codes for a given graph, then $\{\mathcal{A}'_{t}\}$ is also a sequence of codes for the same graph. To the contrary, if $\{\mathcal{A}'_{t}\}$ is a sequence of codes for a graph, then $\{{\mathcal{A}}_n\}$ is called a sequence of \emph{quasi-codes} for the same graph.

Let $\bm{b}_i$, $i=1,2,\ldots,T$ be any string with entries in $\mathcal{X}$. Let $\mathcal{B}_n\triangleq \{\bm{b}_1,\bm{b}_2,\ldots,\bm{b}_T\}^*\cap \mathcal{X}^n$ be uniquely decomposable\footnote{For a set of strings $\{\bm{b}_1,\bm{b}_2,\ldots,\bm{b}_T\}$ defined above, $\{\bm{b}_1,\bm{b}_2,\ldots,\bm{b}_T\}^*$ denotes the family of all sequences which are concatenations of these $b_i$, $i=1,2,..., T$.}, i.e., any sequence in $\mathcal{B}_n$ can be uniquely decomposed into a sequence of strings in $\{\bm{b}_1,\bm{b}_2,\ldots,\bm{b}_T\}$. If there exists a prefix ${\bm{s}}_\mathrm{p}$ and a suffix ${\bm{s}}_\mathrm{s}$ such that by adding a prefix ${\bm{s}}_\mathrm{p}$ and a suffix ${\bm{s}}_\mathrm{s}$ to all sequences in  $\mathcal{B}_n$, we obtain a new set denoted by $\mathcal{B}'_t$. If $\{\mathcal{B}'_t\}$ is a sequence of codes for the graph $G$, then we call $\{{\mathcal{B}}_n\}$ a sequence of \emph{quasi $T$-codes} for $G$. Moreover, if $\{\mathcal{B}_n\}$ is a sequence of codes for $G$, then we call $\{{\mathcal{B}}_n\}$ a sequence of \emph{$T$-codes} for $G$.
For any sequence $\bm{x}$, let $\ell(\bm{x})$ denote the length of $\bm{x}$.
By \cite[Lemma~4.5]{2011Csiszar}, we have
$$R(\{{\mathcal{B}}_n\})=-\log x^*,$$ where $x^*$ is the only positive root of $$\sum_{t=1}^T x^{\ell(\bm{b}_t)}=1.$$
Moreover, if $\{{\mathcal{B}}_n\}$ achieves the largest rate of a sequence of (quasi) $T$-code for $G$, we say that $\{{\mathcal{B}}_n\}$ is an \emph{asymptotically optimal (quasi) $T$-code}.

Now we define two kinds of mappings $T_\mathrm{r}$ and $T_\pi$ as follows.
\begin{itemize}
\item If a graph $G'$  is obtained from a graph $G$ by reversing the sequences representing the vertices of $G$, then $T_\mathrm{r}(G)=G'$.
\item Let $\pi$ be a permutation of $\mathcal{X}$. Let $G''$ be a graph such that for any pair $v,u$, $uv\in E(G)$ if and only if $\pi(u)\pi(v)\in E(G'')$. Then $T_\pi(G)=G''$.
\end{itemize}
Two graphs $G_1$ and $G_2$ are called \emph{interchangeable} if there exists a permutation $\pi$ of $\mathcal{X}$ such that one one of the following three conditions holds.
\begin{enumerate}
\item $T_\mathrm{r}(G_1)=G_2$
\item $T_\pi(G_1)=G_2$
\item $T_\pi(T_\mathrm{r}(G_1))=G_2$
\end{enumerate}
Obviously, the zero-error
capacities of two interchangeable graphs are the same. Table~I lists all the graphs with one edge representing the binary channels with two memories. The graphs in each case are pairwise interchangeable.
We only need to consider any one of the graphs in each case.

\section{Capacity of the Graphs with One Edge}\label{sec:oneedge}
In this section, we first construct an optimal quasi 2-code for any graph with one edge. Then we will show that this quasi 2-code is asymptotically optimal for a class of graphs with one edge.

\subsection{optimal quasi 2-code for the graph with one edge}
Recall that the graph $G$ with only one edge is represented by $G(\bm{u},\bm{v})$, where $\bm{uv}$ is the only edge. Before we construct the optimal (quasi) 2-code for $G(\bm{u},\bm{v})$, we first show the three definitions following.

\begin{defn}
The sequence $\bm{x}$ is a \emph{unit} of $\bm{y}$ if $\ell(\bm{x})\le \ell(\bm{y})$ and there exists a non-negative number $t<\ell(\bm{x})$ such that $y_i=x_{i-t \bmod \ell(\bm{x})}$ for any $i\in \mathbb{Z}[0,\ell(\bm{y})-1]$.
Moreover, $\bm{x}$ is called a \emph{prefix-unit} of $\bm{y}$ if $t=0$ and $\bm{x}$ is called a \emph{suffix-unit} of $\bm{y}$ if $t=\ell(\bm{y}) \bmod \ell(\bm{x})$.
\end{defn}
Clearly, any sequence is a prefix-unit and suffix-unit of itself.
To facilitate the channel capacity characterization, given $G(\bm{u},\bm{v})$ with $\bm{u},\bm{v}\in\mathcal{X}^{m+1}$, we introduce the following notations.
\begin{itemize}
\item Let $\mathrm{pre}(\bm{u},\bm{v})$ and $\mathrm{suf}(\bm{u},\bm{v})$ denote, respectively, the longest common prefix and suffix of $\bm{u}$ and $\bm{v}$;
\item Let $\ell_\mathrm{p}(\bm{u},\bm{v})\triangleq \ell(\mathrm{pre}(\bm{u},\bm{v}))$ and
$\ell_\mathrm{s}(\bm{u},\bm{v})\triangleq \ell(\mathrm{suf}(\bm{u},\bm{v}))$. With a slight abuse of notation, let
$$\ell(\bm{u},\bm{v})\triangleq \max\{\ell_\mathrm{p}(\bm{u},\bm{v}),\ell_\mathrm{s}(\bm{u},\bm{v})\}.$$
Since the zero-error capacities of $G(\bm{u},\bm{v})$ and $T_\mathrm{r}(G(\bm{u},\bm{v}))$ are the same, we only need to consider the case that $\ell_\mathrm{p}(\bm{u},\bm{v})\ge \ell_\mathrm{s}(\bm{u},\bm{v})$, i.e., $\ell(\bm{u},\bm{v})=\ell_\mathrm{p}(\bm{u},\bm{v})$.
\item Let $\bm{u_v}$ denote the shortest prefix-unit of $\bm{u}$ such that $\ell(\bm{u_v})\ge m+1-\ell(\bm{u},\bm{v})$. Likewise, let $\bm{v_u}$ denote the shortest prefix-unit of $\bm{v}$ such that $\ell(\bm{v_u})\ge m+1-\ell(\bm{u},\bm{v})$.
\end{itemize}

Now we have the following theorem.
\begin{theorem}\label{thm:2-code}
For $G(\bm{u},\bm{v})$ with $\bm{u},\bm{v}\in\mathcal{X}^{m+1}$ and $\ell(\bm{u},\bm{v})=\ell_\mathrm{p}(\bm{u},\bm{v})$, the set $\mathcal{B}_n\triangleq\{\bm{u_v},\bm{v_u}\}^*\cap \mathcal{X}^n$ is an asymptotically optimal quasi 2-code. Moreover, $C(G(\bm{u},\bm{v}))\ge -\log x^*$, where $x^*$ is the only positive root of
$$x^{\ell(\bm{u_v})}+x^{\ell(\bm{v_u})}=1.$$
\end{theorem}
The following two lemmas will serve as stepping stones to establish Theorem~\ref{thm:2-code}.
\begin{lemma}\label{lm:2-code}
For $\bm{u},\bm{v}\in\mathcal{X}^{m+1}$, let $\bm{x}\in \{\bm{u_v},\bm{v_u}\}^*$ be an arbitrary but fixed sequence.
Let $\bm{y}=\bm{x}\mathrm{pre}(\bm{u},\bm{v})$, the concatenation of $\bm{x}$ and $\mathrm{pre}(\bm{u},\bm{v})$.
If $\ell(\bm{u},\bm{v})>0$, then for any $i\in \mathbb{Z}[0,\ell(\bm{u},\bm{v})-1]$, we have	$y_i=u_i=v_i$.
\end{lemma}
\begin{IEEEproof}
The sequence $\bm{y}$ can be written as $\bm{x}^{(0)}\bm{x}^{(1)} \ldots \bm{x}^{(N-1)}\bm{x}^{(N)}$, where $\bm{x}^{(n)}\in \{\bm{u_v},\bm{v_u}\}$, $n\in \mathbb{Z}[0,N-1]$, and $\bm{x}^{(N)}=\mathrm{pre}(\bm{u},\bm{v})$. Now we prove that $y_i=u_i=v_i$ for any $i\in \mathbb{Z}[0,\ell(\bm{u},\bm{v})-1]$.
Note for each $y_i$, $i\in \mathbb{Z}[0,\ell(\bm{u,v})-1]$, there exists a string $\bm{x}^{(n_i)}$, $n_i\in \mathbb{Z}[0,N]$ such that $y_i=x^{(n_i)}_{\tilde{i}}$, an entry of $\bm{x}^{(n_i)}$.
Note that
$$\sum_{j=0}^{n_i-1}\ell(\bm{x}_j)\le i \text{ and } \sum_{j=0}^{n_i}\ell(\bm{x}_j)> i.$$
We have
\begin{equation*}\label{eq:lm2-1}
    n_i= \max\left\{n\in \mathbb{Z}:\sum_{j=0}^{n-1}\ell(\bm{x}_j)\le  i\right\}
\end{equation*}
and
$$\tilde{i}= i-\sum_{j=0}^{n_i-1}\ell(\bm{x}_j).$$
Now we denote
$$n_i'\triangleq \sum_{j=0}^{n_i-1} \mathbbm{1}(\bm{x}_j=\bm{u_v})$$
and
$$n_i''\triangleq \sum_{j=0}^{n_i-1} \mathbbm{1}(\bm{x}_j=\bm{v_u}),$$
where $\mathbbm{1}(\cdot)$ is the indicator function.
Clearly, $i=n_i' \ell(\bm{u_v})+n_i'' \ell(\bm{v_u})+\tilde{i}$.

By the definition of $\bm{u_v}$ and $\bm{v_u}$, for any $j\in \mathbb{Z}[0,\ell(\bm{u},\bm{v})-1]$, we have
\begin{equation}\label{eq:ujvj}
    u_j=u_{j\bmod \ell(\bm{u_v})}=v_j=v_{j\bmod \ell(\bm{v_u})},
\end{equation}
and for any $j_1,j_2\in  \mathbb{Z}[0,\ell(\bm{u},\bm{v})-1]$, we have
\begin{enumerate}
\item $j_1 \equiv j_2 \bmod \ell(\bm{u_v})$ implies $u_{j_1}=u_{j_2}$
\item $j_1 \equiv j_2 \bmod \ell(\bm{v_u})$ implies $v_{j_1}=v_{j_2}$.
\end{enumerate}
When $\bm{x}_{n_i}=\bm{u_v}$, and we have
$${y}_{i}=x^{(n_i)}_{\tilde{i}}=(\bm{u_v})_{\tilde{i}}=u_{\tilde{i}}\stackrel{(a)}{=}u_{\tilde{i}+n_i' \ell(\bm{u_v})}\stackrel{(b)}{=}v_{\tilde{i}+n_i' \ell(\bm{u_v})}\stackrel{(c)}{=}v_{\tilde{i}+n_i' \ell(\bm{u_v})+n_i'' \ell(\bm{v_u})}=v_i=u_i,$$
where $(a)$ and $(c)$ hold since Condition 1 and 2 above hold, and $(b)$ holds since \eqref{eq:ujvj} holds.

Likewise, when $\bm{x}_{n_i}=\bm{v_u}$. We have
$${y}_{i}=x^{(n_i)}_{\tilde{i}}=(\bm{v_u})_{\tilde{i}}=v_{\tilde{i}}=v_{\tilde{i}+n_i'' \ell(\bm{v_u})}=u_{\tilde{i}+n_i'' \ell(\bm{v_u})}=u_{\tilde{i}+n_i' \ell(\bm{u_v})+n_i'' \ell(\bm{v_u})}=u_i=v_i.$$
When $\bm{x}_{n_i}=\mathrm{pre}(\bm{u},\bm{v})$, we have
$x^{(n_i)}_{\tilde{i}}=u_{\tilde{i}}=v_{\tilde{i}}$,
and thus,
$${y}_{i}=x^{(n_i)}_{\tilde{i}}=v_{\tilde{i}}=v_{\tilde{i}+n_i'' \ell(\bm{v_u})}=u_{\tilde{i}+n_i'' \ell(\bm{v_u})}=u_{\tilde{i}+n_i' \ell(\bm{u_v})+n_i'' \ell(\bm{v_u})}=u_i=v_i.$$
\end{IEEEproof}

\begin{example}
For $G=G(001001,001000)$, we have $\mathrm{pre}(\bm{u},\bm{v})=00100$ and $\ell(\bm{u},\bm{v})=5$.
Then $\bm{u_v}=001$ and $\bm{v_u}=0010$. Let $\bm{y}$ be an arbitrary but fixed sequence in $\{\bm{u_v},\bm{v_u}\}^*\times \{\mathrm{pre}(\bm{u},\bm{v})\}=\{001,0010\}^*\times \{00100\}$. By Lemma~\ref{lm:2-code}, we have $y_i=u_i=v_i$ for any $i\in \mathbb{Z}[0,4]$, i.e.,
$y_0y_1y_2y_3y_4=00100.$
\end{example}

\begin{lemma}\label{lm:2-code-2}
Let $\mathcal{C}_n=\{\bm{c}_1,\bm{c}_2,\cdots,\bm{c}_T\}^*\cap \mathcal{X}^n$ for each $n$, where $\bm{c}_i$, $i=1,2,...,T$, are strings with entries in $\mathcal{X}$.
For $G(\bm{u},\bm{v})$ with $\bm{u},\bm{v}\in\mathcal{X}^{m+1}$ and $\ell(\bm{u},\bm{v})=\ell_\mathrm{p}(\bm{u},\bm{v})$, if $\{\mathcal{C}_n\}$ is a sequence of quasi codes, then any $\bm{x}\in \{\bm{c}_1,\bm{c}_2,\cdots,\bm{c}_T\}$ with $\ell(\bm{x})\le m+1$ is a unit of $\bm{u}$ or $\bm{v}$.
Moreover, if $\bm{x}$ is a unit of $\bm{u}$, then $\ell(\bm{x})\ge \ell(\bm{u_v})$.
\end{lemma}
\begin{IEEEproof}
Note that $\{\mathcal{C}_n\}$ is a sequence of quasi codes. We can find a prefix ${\bm{s}}_\mathrm{p}$ and a suffix ${\bm{s}}_\mathrm{s}$ such that for any $n$, by adding ${\bm{s}}_\mathrm{p}$ and ${\bm{s}}_\mathrm{s}$ to all sequences in $\mathcal{C}_n$, we obtain a code $\mathcal{C}'_{n'}$ for $G(\bm{u},\bm{v})$ with $n'=\ell({\bm{s}}_\mathrm{p})+n+\ell({\bm{s}}_\mathrm{s})$. Let $\{\mathcal{C}'_{n'}\}$ be a sequence of these codes indexed by $n'$.
For any sequence $\bm{s}$ and positive integer $w$, let $\bm{s}^w$ denote the concatenation of $\bm{s}$ with itself $w$ times.

Now we consider $n'= \ell({\bm{s}}_\mathrm{p})+\left(2m+\ell(\bm{y})\right)\ell(\bm{x})+\ell({\bm{s}}_\mathrm{s})$. Let
$$\bm{c}'\triangleq {\bm{s}}_\mathrm{p} \bm{x}^{2m+\ell(\bm{y})} {\bm{s}}_\mathrm{s} \text{ and }
\bm{c}''\triangleq {\bm{s}}_\mathrm{p} \bm{x}^{m} \bm{y}^{\ell(\bm{x})} \bm{x}^{m} {\bm{s}}_\mathrm{s}$$
be two codewords in $\mathcal{C}'_{n'}$,
where $\bm{y}$ is an arbitrary string in $\{\bm{c}_1,\bm{c}_2,\cdots,\bm{c}_T\}$ with $\bm{y}\neq \bm{x}$.
Since $\bm{c}'$ and $\bm{c}''$ are distinguishable for $G(\bm{u},\bm{v})$, there exists a coordinate $i\in \mathbb{Z}[0,n'-m-1]$ such that $$\{c'_ic'_{i+1}\cdots c'_{i+m}, c''_ic''_{i+1}\cdots c''_{i+m}\}=\{\bm{u}, \bm{v}\}.$$
Without loss of generality, we assume $\bm{u}=c'_ic'_{i+1}\cdots c'_{i+m}$ and $\bm{v}=c''_ic''_{i+1}\cdots c''_{i+m}$, and we will prove that $\bm{x}$ is a unit of $\bm{u}$.

By definition, $u_{\ell(\bm{u},\bm{v})}\neq v_{\ell(\bm{u},\bm{v})}$, which implies that
$c'_{i+\ell(\bm{u},\bm{v})}\neq c''_{i+\ell(\bm{u},\bm{v})}$.
On the other hand, the first $\ell({\bm{s}}_\mathrm{p})+m\times\ell(\bm{x})$ and the last $m\times\ell(\bm{x})+\ell({\bm{s}}_\mathrm{s})$ bits of $\bm{c}'$ and $\bm{c}''$ are respectively the same, i.e., $c'_j= c''_j$ for $j\in \mathbb{Z}[0,\ell({\bm{s}}_\mathrm{p})+m\times\ell(\bm{x})-1]\cup \mathbb{Z}[n'-m\times\ell(\bm{x})-\ell({\bm{s}}_\mathrm{s}),n'-1]$. Therefore,
$$i+\ell(\bm{u},\bm{v}) \in \mathbb{Z}[\ell({\bm{s}}_\mathrm{p})+m\times\ell(\bm{x}), n'-m\times\ell(\bm{x})-\ell({\bm{s}}_\mathrm{s})-1].$$
Thus,
$$i\ge \ell({\bm{s}}_\mathrm{p})+m\times\ell(\bm{x})-\ell(\bm{u},\bm{v}) \ge \ell({\bm{s}}_\mathrm{p})+m-m = \ell({\bm{s}}_\mathrm{p})$$
 and
$$i+m\le n'-m\times\ell(\bm{x})-\ell({\bm{s}}_\mathrm{s})-1+m-\ell(\bm{u},\bm{v})\le n'-\ell({\bm{s}}_\mathrm{s})-1.$$
Recall that $\bm{c}'= {\bm{s}}_\mathrm{p} \bm{x}^{2m+\ell(\bm{y})} {\bm{s}}_\mathrm{s}$. Letting $\oplus$ denote the modulo-$\ell(\bm{x})$ addition and $i'= i\oplus\left(-\ell(\bm{s}_\mathrm{p})\right) $, we have
$$\bm{u}= c'_ic'_{i+1}\cdots c'_{i+m}=x_{i'}x_{i'\oplus 1}\cdots x_{i'\oplus m}.$$
Note that $\ell(\bm{x})\le m+1$. We have
$\bm{x}'\triangleq x_{i'}x_{{i'\oplus 1}}\cdots x_{i'\oplus (\ell(\bm{x})-1)}$ is a prefix-unit of $\bm{u}$, and then
$\bm{x}$ is a unit of $\bm{u}$.

Now we prove that if $\bm{x}$ is a unit of $\bm{u}$, then $\ell(\bm{x})\ge\ell(\bm{u_v})$. Assume the contrary, $\ell(\bm{x})<\ell(\bm{u_v})$.
Recall that $\bm{u_v}$ is the shortest prefix-unit of $\bm{u}$ such that $\ell(\bm{u_v})\ge m+1-\ell(\bm{u},\bm{v})$. As a prefix shorter than $\bm{u_v}$, we have $\ell(\bm{x})<m+1-\ell(\bm{u},\bm{v})\le m+1-\ell_\mathrm{s}(\bm{u},\bm{v})$, which implies $\ell(\bm{u},\bm{v})+\ell(\bm{x})<m+1$ and  $\ell(\bm{u},\bm{v})+\ell_\mathrm{s}(\bm{x})<m+1$.
Note that $\ell(\bm{u})=m+1$. Now for $u_{\ell(\bm{u},\bm{v})+\ell(\bm{x})}$ and $u_{m-\ell_\mathrm{s}(\bm{u},\bm{v})-\ell(\bm{x})}$, as $\bm{x}$ is a unit of $\bm{u}$,
\begin{equation}\label{eq:lm-3}        u_{\ell(\bm{u},\bm{v})+\ell(\bm{x})}=u_{\ell(\bm{u},\bm{v})}
\text{ and }
    u_{m-\ell_\mathrm{s}(\bm{u},\bm{v})-\ell(\bm{x})}=u_{m-\ell_\mathrm{s}(\bm{u},\bm{v})}.
\end{equation}

Now we consider $n''=\ell({\bm{s}}_\mathrm{s})+(2m+1)\ell(\bm{x})+m\ell(\bm{y})+\ell({\bm{s}}_\mathrm{s})$. Let
\begin{equation}\label{eq:lm-4}
\bm{d}'\triangleq {\bm{s}}_\mathrm{p} \bm{x}^{m+1} \bm{y}^{m} \bm{x}^m {\bm{s}}_\mathrm{s}
\text{ and }
\bm{d}''\triangleq {\bm{s}}_\mathrm{p} \bm{x}^m \bm{y}^{m} \bm{x}^{m+1} {\bm{s}}_\mathrm{s}
\end{equation}
be two codewords in $\mathcal{C}'_{n'}$. There exists a coordinate $i\in \mathbb{Z}[0,n''-m-1]$ such that one of the following two conditions holds.
\begin{enumerate}
    \item $d'_id'_{i+1}\cdots d'_{i+m}=\bm{u}$ and $d''_id''_{i+1}\cdots d''_{i+m}=\bm{v}$;
    \item $d'_id'_{i+1}\cdots d'_{i+m}=\bm{v}$ and $d''_id''_{i+1}\cdots d''_{i+m}=\bm{u}$.
\end{enumerate}
Note that $u_{\ell(\bm{u},\bm{v})} \neq v_{\ell(\bm{u},\bm{v})}$ and $u_{m-\ell_\mathrm{s}(\bm{u},\bm{v})} \neq v_{m-\ell_\mathrm{s}(\bm{u},\bm{v})}$. We have
\begin{equation}\label{eq:lm-5}
d'_{i_1}\neq d''_{i_1} \text{ and } d'_{i_2}\neq d''_{i_2},
\end{equation}
where $i_1=i+\ell(\bm{u},\bm{v})$ and $i_2=i+m-\ell_\mathrm{s}(\bm{u},\bm{v})$.
To simplify the notations, let $I_1\triangleq \ell({\bm{s}}_\mathrm{s})+m\ell(\bm{x})$, $I_2\triangleq I_1+\ell(\bm{x})$, $I_3\triangleq I_1+m\ell(\bm{y})$ and $I_4\triangleq I_1+\ell(\bm{x})+m\ell(\bm{y})$. Clearly, $i_1\le i_2$ and $I_1<\min\{I_2,I_3\}\le \max \{I_2,I_3\}<I_4$. If $i_1\in \mathbb{Z}[0,I_1-1]\cup \mathbb{Z}[I_4,n''-1]$, then $d'_{i_1}= d''_{i_1}$ which contradicts (\ref{eq:lm-5}). Thus $i_1\in \mathbb{Z}[I_1,I_4-1].$ Likewise, we also have $i_2\in \mathbb{Z}[I_1,I_4-1].$

We first assume that Condition 1) holds. If $i_1\in \mathbb{Z}[I_1,I_3-1]$, then by (\ref{eq:lm-3}), we have
$$d'_{i_1}=u_{\ell(\bm{u},\bm{v})}=u_{\ell(\bm{u},\bm{v})+\ell(\bm{x})}=d'_{i_1+\ell(\bm{x})}.$$
Thus
$$d'_{i_1}=d'_{i_1+\ell(\bm{x})}=(\bm{s}_\mathrm{p} \bm{x}^{m+1} \bm{y}^{m})_{i_1+\ell(\bm{x})}=(\bm{y}^m)_{i_1+\ell(\bm{x})-I_2}=(\bm{y}^m)_{i_1-I_1}=(\bm{s}_\mathrm{p} \bm{x}^m \bm{y}^{m})_{i_1}=d''_{i_1},$$
which contradicts (\ref{eq:lm-5}). Therefore, $i_1\in \mathbb{Z}[I_3,I_4-1]$. Recall that $i_2\ge i_1$ and $i_2\in \mathbb{Z}[I_1,I_4-1]$. We have $i_2\in \mathbb{Z}[I_3,I_4-1]$, and so
$$u_{m-\ell_\mathrm{s}(\bm{u},\bm{v})}=d'_{i_2}=d'_{i_2-\ell(\bm{x})}=d''_{i_2},$$ which contradicts (\ref{eq:lm-5}). Hence, there do not exist coordinates $i_1$ and $i_2$ such that Condition 1 holds.

Now we consider that Condition 2 holds. If $i_2\in \mathbb{Z}[I_2,I_4-1]$, likewise, we can obtain that
$$d''_{i_2}=u_{m-\ell_\mathrm{s}(\bm{u},\bm{v})}=u_{m-\ell_\mathrm{s}(\bm{u},\bm{v})-\ell(\bm{x})}=d''_{i_2-\ell(\bm{x})}=(\bm{s}_\mathrm{p} \bm{x}^m\bm{y}^{m}\bm{x})_{i_1-\ell(\bm{x})}=(\bm{s}_\mathrm{p} \bm{x}^{m+1}\bm{y}^{m})_{i_1}=d'_{i_2},$$
which contradicts (\ref{eq:lm-5}). Therefore, $i_2\in \mathbb{Z}[I_1,I_2-1]$, and then $i_1\in \mathbb{Z}[I_1,I_2-1]$, and so
$$u_{\ell(\bm{u},\bm{v})}=d''_{i_2}=d''_{i_2-\ell(\bm{x})}=d'_{i_2},$$ which contradicts (\ref{eq:lm-5}). Hence, there do not exist coordinates $i_1$ and $i_2$ such that Condition 2 holds. Therefore, $\ell(\bm{x})\ge \ell(\bm{u_v})$ if $\bm{x}$ is a unit of $\bm{u}$.
\end{IEEEproof}

\begin{corollary}
Let $\{\mathcal{C}_n\}$ be a sequence of quasi codes for $G(\bm{u},\bm{v})$ with $\mathcal{C}_n\triangleq\{\bm{c}_1,\bm{c}_2,\cdots,\bm{c}_T\}^*\cap \mathcal{X}^n$. Then $\ell(\bm{c}_t)\ge \min \{\ell(\bm{u_v}),\ell(\bm{v_u})\}$, $t=1,2,...,T$.
\end{corollary}

With the above auxiliary results, we turn to the proof of Theorem~\ref{thm:2-code}.
\begin{IEEEproof}[Proof of Theorem~\ref{thm:2-code}]	
We first prove that $\mathcal{B}_n$ is a quasi 2-code for $G(\bm{u},\bm{v})$. It is sufficient to prove that
$$\mathcal{B}'_{n'}\triangleq\left(\{\bm{u_v},\bm{v_u}\}^*\cap \mathcal{X}^n\right)\times\{\mathrm{pre}(\bm{u},\bm{v})\}$$
is a code for $G(\bm{u},\bm{v})$ with $n'\triangleq n+\ell_\mathrm{p}(\bm{u},\bm{v})$, i.e.,
any two different sequences $\bm{b}',\bm{b}''\in \mathcal{B}'_n$ are distinguishable for $G(\bm{u},\bm{v})$.	
The sequences $\bm{b}',\bm{b}''\in \mathcal{B}'_n$ can be written as $\bm{b}'=\bm{b}'_{0}\bm{b}'_{1}\ldots \bm{b}'_{N_1-1}\bm{b}'_{N_1}$ and $\bm{b}''=\bm{b}''_{0}\bm{b}''_{1}\ldots \bm{b}''_{N_2-1}\bm{b}''_{N_2}$, where $\bm{b}'_{n_1},\bm{b}''_{n_2}\in \{\bm{u_v},\bm{v_u}\}$ for $n_1\in \mathbb{Z}[0,N_1-1]$ and $n_2\in \mathbb{Z}[0,N_2-1]$, and $\bm{b}'_{N_1}=\bm{b}''_{N_2}=\mathrm{pre}(\bm{u},\bm{v})$.
Let $n^*$ be the smallest index such that $\bm{b}'_{n^*}\neq\bm{b}''_{n^*}$, i.e.,
$$n^*\triangleq \min\left\{n\in \mathbb{Z}[0,\min\{N_1,N_2\}-1]: \bm{b}'_{n}\neq\bm{b}''_{n}\right\}.$$
Without loss of generality, we assume that $\bm{b}'_{n^*}=\bm{{u}_{v}}$ and $\bm{b}''_{n^*}=\bm{{v}_{u}}$. Let
$$i\triangleq \sum_{n=0}^{n^*-1} \ell(\bm{b}'_n)= \sum_{n=0}^{n^*-1} \ell(\bm{b}''_n)$$
be the coordinate of the first symbol of $\bm{b}'_{n^*}$ in $\bm{b}'$.

Note that both $\bm{b}'_{n^*+1}\bm{b}'_{n^*+2}\ldots \bm{b}'_{N_1-1}$ and $\bm{b}''_{n^*+1}\bm{b}''_{n^*+2}\ldots \bm{b}''_{N_2-1}$ are in $\{\bm{u_v},\bm{v_u}\}^*$.
By Lemma~\ref{lm:2-code}, we have
\begin{equation}\label{eq:thm1-1}
b'_{i+\ell(\bm{u_v})+j}=b''_{i+\ell(\bm{v_u})+j}=u_j=v_j,\forall j\in \mathbb{Z}[0,\ell(\bm{u,v})-1].
\end{equation}
On the other hand, the fact that $\bm{u_v}$ and $\bm{v_u}$ are, respectively, the prefix-unit of $\bm{u}$ and $\bm{v}$ implies that
\begin{equation}\label{eq:thm1-21}
u_j=u_{j+\ell(\bm{u_v})}, \forall j\in \mathbb{Z}[0,m-\ell(\bm{u_v})]
\end{equation}
and
\begin{equation}\label{eq:thm1-22}
v_j=v_{j+\ell(\bm{v_u})}, \forall j\in \mathbb{Z}[0,m-\ell(\bm{v_u})].
\end{equation}
Recall that $\ell(\bm{u_v}),\ell(\bm{v_u})\ge m+1-\ell(\bm{u},\bm{v})$. By (\ref{eq:thm1-1}) and (\ref{eq:thm1-21}), we have
$$b'_{i+\ell(\bm{u_v})+j}=u_{\ell(\bm{u_v})+j}, \forall j\in \mathbb{Z}[0,\ell(\bm{u,v})]\cap \mathbb{Z}[0,m-\ell(\bm{u_v})]=\mathbb{Z}[0,m-\ell(\bm{u_v})].$$
Therefore,
\begin{equation*}
    \begin{split}
        &b'_{i}b'_{i+1}\cdots b'_{i+m}\\
        =&b'_{i}b'_{i+1}\cdots b'_{i+\ell(\bm{u_v})-1}b'_{i+\ell(\bm{u_v})}\cdots b'_{i+m}\\
        =&\bm{b}'_{n^*} b'_{i+\ell(\bm{u_v})}\cdots b'_{i+m}\\
        =&\bm{u_v} u_{\ell(\bm{u_v})}\cdots u_{m}\\
        =& \bm{u}.
    \end{split}
\end{equation*}
Likewise, we can also obtain that
\begin{equation*}
    b''_{i}b''_{i+1}\cdots b''_{i+m}=\bm{v}.
\end{equation*}
Hence $\bm{b}',\bm{b}''\in \mathcal{B}'_n$ are distinguishable for $G(\bm{u},\bm{v})$, which indicates that $\mathcal{B}_n$ is a quasi 2-code for $G(\bm{u},\bm{v})$.

We now prove that $\{\mathcal{B}_n\}$ has the highest rate among all the quasi 2-codes for $G(\bm{u},\bm{v})$.
Assume there exists a quasi 2-code $$\mathcal{C}_n\triangleq \{\bm{x},\bm{y}\}^*\cap \mathcal{X}^n$$ such that $R(\{\mathcal{C}_n\})>R(\{\mathcal{B}_n\}).$
Note that $R(\{\mathcal{B}_n\})=-\log x^*$ and $R(\{\mathcal{C}_n\})=-\log x'$, where $x^*$ and $x'$ satisfy, respectively, that
$$(x^*)^{\ell(\bm{u_v})}+(x^*)^{\ell(\bm{v_u})}=1$$ and $$(x')^{\ell(\bm{x})}+(x')^{\ell(\bm{y})}=1.$$
If $\ell(\bm{x})\ge\ell(\bm{u_v})$ and $\ell(\bm{y})\ge\ell(\bm{v_u})$, then $x^*\ge x'$, which contradicts the assumption that $R(\{\mathcal{B}_n\})<R(\{\mathcal{C}_n\}).$
Thus, $\ell(\bm{x})<\ell(\bm{u_v})\le m+1$ or $\ell(\bm{y})<\ell(\bm{v_u})\le m+1$. By Lemma~\ref{lm:2-code-2}, either $\bm{x}$ or $\bm{y}$ is a unit of $\bm{u}$ or $\bm{v}$.
Without loss of generality, let $\bm{x}$ be a unit of $\bm{u}$ and $\ell(\bm{x})\ge \ell(\bm{u_v})$.
Moreover, since $R(\{\mathcal{C}_n\})>R(\{\mathcal{B}_n\})$, we have $\ell(\bm{y}) < \ell(\bm{v_u})\le m+1$. Thus, by Lemma~\ref{lm:2-code-2}, $\bm{y}$ is also a unit of $\bm{u}$ or $\bm{v}$. Considering that $\ell(\bm{y}) < \ell(\bm{v_u})$, we can obtain that $\bm{y}$ is a unit of $\bm{u}$ but not a unit of $\bm{v}$, and thus $\ell(\bm{u_v})\le \ell(\bm{y}) <  \ell(\bm{v_u})$. Moreover, we can also obtain that $\bm{x}$ is not a unit of $\bm{v}$.

We have proved that for $\{\mathcal{C}_n\}$, each of $\bm{x}$ and $\bm{y}$ is a unit of $\bm{u}$, and neither of them is a unit of $\bm{v}$. The remaining part will prove that this statement does not hold, so that there does not exist a sequence of codes  $\{\mathcal{C}_n\}$ whose rate would be larger than $R(\{\mathcal{B}_n\})$. This will complete the proof of this theorem.

Now we assume that the statement holds. There exists a prefix ${\bm{s}}_\mathrm{p}$ and a suffix ${\bm{s}}_\mathrm{s}$ such that by adding ${\bm{s}}_\mathrm{p}$ and ${\bm{s}}_\mathrm{s}$ to all sequences in  $\mathcal{C}_n$, we obtain a code for $G(\bm{u},\bm{v})$, denoted by $$\mathcal{C}'_t\triangleq \left(\{{\bm{s}}_\mathrm{p}\}\times \{\bm{x},\bm{y}\}^*\times \{{\bm{s}}_\mathrm{s}\}\right)\cap \mathcal{X}^t.$$

Let
$$\bm{c}^{xxx}\triangleq {\bm{s}}_\mathrm{p} \bm{x}^{m\ell(\bm{y})} \bm{x}^{m\ell(\bm{y})} \bm{x}^{m\ell(\bm{y})} {\bm{s}}_\mathrm{s}$$
and
$$\bm{c}^{xyx}\triangleq {\bm{s}}_\mathrm{p} \bm{x}^{m\ell(\bm{y})} \bm{y}^{m\ell(\bm{x})} \bm{x}^{m\ell(\bm{y})} {\bm{s}}_\mathrm{s}$$
be sequences of length
$t'\triangleq\ell({\bm{s}}_\mathrm{p})+3m\ell(\bm{x})\ell(\bm{y})+\ell({\bm{s}}_\mathrm{s})$.
Note that these two sequences are distinguishable.
There exists a coordinate $i$ such that
$$\{c^{xxx}_ic^{xxx}_{i+1}\cdots c^{xxx}_{i+m}, c^{xyx}_ic^{xyx}_{i+1}\cdots c^{xyx}_{i+m}\}=\{\bm{u}, \bm{v}\}.$$
Letting $i_\mathrm{p}\triangleq i+\ell(\bm{u},\bm{v})$ and $i_\mathrm{s}\triangleq i+m-\ell_\mathrm{s}(\bm{u},\bm{v})$, since $c^{xxx}_{i_\mathrm{p}}\neq c^{xyx}_{i_\mathrm{p}}$ and $c^{xxx}_{i_\mathrm{s}}\neq c^{xyx}_{i_\mathrm{s}}$, we have
$$\ell\left({\bm{s}}_\mathrm{p} \bm{x}^{m\ell(\bm{y})} \right)\le i_\mathrm{p}\le i_\mathrm{s}< \ell\left({\bm{s}}_\mathrm{p} \bm{x}^{m\ell(\bm{y})} \bm{y}^{m\ell(\bm{x})}\right),$$
which implies
\begin{equation}\label{eq:thm1-6}
i\in \mathbb{Z}\left[\ell\left({\bm{s}}_\mathrm{p} \bm{x}^{m\ell(\bm{y})} \right)-\ell(\bm{u},\bm{v}), \ell\left({\bm{s}}_\mathrm{p} \bm{x}^{m\ell(\bm{y})} \bm{y}^{m\ell(\bm{x})}\right)-m+\ell_\mathrm{s}(\bm{u},\bm{v})-1\right].
\end{equation}
On the other hand, as $\bm{x}$ is not a unit of $\bm{v}$, we have $c^{xxx}_ic^{xxx}_{i+1}\cdots c^{xxx}_{i+m}=\bm{u}$ and thus $c^{xyx}_ic^{xyx}_{i+1}\cdots c^{xyx}_{i+m}=\bm{v}$.
Since $\bm{y}$ is neither a unit of $\bm{v}$, we further have 
\begin{equation}\label{eqi2}
i\notin \mathbb{Z}\left[\ell\left({\bm{s}}_\mathrm{p} \bm{x}^{m\ell(\bm{y})}\right), \ell\left({\bm{s}}_\mathrm{p} \bm{x}^{m\ell(\bm{y})} \bm{y}^{m\ell(\bm{x})} \right)-m-1\right].
\end{equation}
By \eqref{eq:thm1-6} and \eqref{eqi2}, we can obtain that
$$i\in \mathcal{I}_1\triangleq\mathbb{Z}\left[\ell\left({\bm{s}}_\mathrm{p} \bm{x}^{m\ell(\bm{y})} \right)-\ell(\bm{u},\bm{v}), \ell\left({\bm{s}}_\mathrm{p} \bm{x}^{m\ell(\bm{y})} \right)-1\right]$$
or
$$i\in \mathcal{I}_2\triangleq\mathbb{Z}\left[\ell\left({\bm{s}}_\mathrm{p} \bm{x}^{m\ell(\bm{y})} \bm{y}^{m\ell(\bm{x})}\right)-m, \ell\left({\bm{s}}_\mathrm{p} \bm{x}^{m\ell(\bm{y})} \bm{y}^{m\ell(\bm{x})}\right)-m+\ell_\mathrm{s}(\bm{u},\bm{v})-1\right].$$
Letting $I_1\triangleq\min\{j\ge 0: x_{j\bmod \ell(\bm{x})}\neq y_{j\bmod \ell(\bm{y})}\}$, if $i\in \mathcal{I}_1$, then $I_1\le \ell(\bm{u,v})$ and
\begin{equation*}
\begin{split}
i_\mathrm{p}=&\min\{j\ge \ell\left({\bm{s}}_\mathrm{p} \bm{x}^{m\ell(\bm{y})}\right):c^{xxx}_{j}\neq c^{xyx}_{j}\}\\
=&\ell\left({\bm{s}}_\mathrm{p} \bm{x}^{m\ell(\bm{y})}\right)+I_1.
\end{split}	
\end{equation*}
Likewise, letting $I_2\triangleq\min\{j> 0: x_{-j\bmod \ell(\bm{x})}\neq y_{-j\bmod \ell(\bm{y})}\}$, if $i\in \mathcal{I}_2$, then $I_2\le \ell_\mathrm{s}(\bm{u,v})$ and
\begin{equation*}
\begin{split}
    i_\mathrm{s}=&\max\{j< \ell\left({\bm{s}}_\mathrm{p} \bm{x}^{m\ell(\bm{y})} \bm{y}^{m\ell(\bm{x})}\right):c^{xxx}_{j}\neq c^{xyx}_{j}\}\\
    =&\ell\left({\bm{s}}_\mathrm{p} \bm{x}^{m\ell(\bm{y})}\bm{y}^{m\ell(\bm{x})}\right)-I_2.
\end{split}	
\end{equation*}
Consider the four cases:
\begin{enumerate}
\item[$A$)] $\bm{c}_\mathrm{p}^{\bm{xx}}=\bm{u}$ and $\bm{c}_\mathrm{p}^{\bm{xy}}=\bm{v}$,
\item[$B$)] $\bm{c}_\mathrm{p}^{\bm{yy}}=\bm{u}$ and $\bm{c}_\mathrm{p}^{\bm{yx}}=\bm{v}$,
\item[$C$)] $\bm{c}_\mathrm{s}^{\bm{yy}}=\bm{u}$ and $\bm{c}_\mathrm{s}^{\bm{xy}}=\bm{v}$,
\item[$D$)] $\bm{c}_\mathrm{s}^{\bm{xx}}=\bm{u}$ and $\bm{c}_\mathrm{s}^{\bm{yx}}=\bm{v}$,
\end{enumerate}
where
$$\bm{c}_\mathrm{p}^{\bm{w}'\bm{w}''} = 
w'_{I_1-\ell(\bm{u,v}) \bmod \ell(\bm{w}')}  ...
w'_{-1 \bmod \ell(\bm{w}')}
w''_{0 \bmod \ell(\bm{w}'')}...
w''_{I_1-\ell(\bm{u,v})+m \bmod \ell(\bm{w}'')},$$
and 
$$\bm{c}_\mathrm{s}^{\bm{w}'\bm{w}''} = 
w'_{-I_2+\ell_\mathrm{s}(\bm{u,v}) -m \bmod \ell(\bm{w}')} ...
w'_{-1 \bmod \ell(\bm{w}')}
w''_{0 \bmod \ell(\bm{w}'')}...
w''_{-I_2+\ell_\mathrm{s}(\bm{u,v}) \bmod \ell(\bm{w}'')},$$
for $\bm{w}',\bm{w}''\in \{\bm{x},\bm{y}\}$.

We have shown that
the fact that $\bm{c}^{xxx}$ and $\bm{c}^{xyx}$ are distinguishable implies that Case $A$ or $D$ holds.
Likewise, we have the following implications.
\begin{itemize}
    \item $\bm{c}^{xxy}\triangleq {\bm{s}}_\mathrm{p} \bm{x}^{m\ell(\bm{y})} \bm{x}^{m\ell(\bm{y})} \bm{y}^{m\ell(\bm{x})} {\bm{s}}_\mathrm{s}$ and $\bm{c}^{xyy}\triangleq {\bm{s}}_\mathrm{p} \bm{x}^{m\ell(\bm{y})} \bm{y}^{m\ell(\bm{x})} \bm{y}^{m\ell(\bm{x})} {\bm{s}}_\mathrm{s}$ are distinguishable $\implies$ Case $A$ or $C$ holds;
    \item $\bm{c}^{yxx}\triangleq {\bm{s}}_\mathrm{p} \bm{y}^{m\ell(\bm{x})} \bm{x}^{m\ell(\bm{y})} \bm{x}^{m\ell(\bm{y})} {\bm{s}}_\mathrm{s}$ and $\bm{c}^{yyx}\triangleq {\bm{s}}_\mathrm{p} \bm{y}^{m\ell(\bm{x})} \bm{y}^{m\ell(\bm{x})} \bm{x}^{m\ell(\bm{y})} {\bm{s}}_\mathrm{s}$ are distinguishable $\implies$ Case $B$ or $D$ holds;
    \item $\bm{c}^{yxy}\triangleq {\bm{s}}_\mathrm{p} \bm{y}^{m\ell(\bm{x})} \bm{x}^{m\ell(\bm{y})} \bm{y}^{m\ell(\bm{x})} {\bm{s}}_\mathrm{s}$ and $\bm{c}^{yyy}\triangleq {\bm{s}}_\mathrm{p} \bm{y}^{m\ell(\bm{x})} \bm{y}^{m\ell(\bm{x})} \bm{y}^{m\ell(\bm{x})} {\bm{s}}_\mathrm{s}$ are distinguishable $\implies$ Case $B$ or $C$ holds.
\end{itemize}
From all these implications, we derive that either both Cases $A$ and $B$ hold, or both Cases $C$ and $D$ hold.
If Cases $A$ and $B$ hold, then
$\bm{c}_\mathrm{p}^{\bm{xx}}=\bm{c}_\mathrm{p}^{\bm{yy}}$, and thus $\bm{c}_\mathrm{p}^{\bm{xx}}=\bm{c}_\mathrm{p}^{\bm{yy}}=\bm{c}_\mathrm{p}^{\bm{yx}}=\bm{c}_\mathrm{p}^{\bm{xy}}$. Hence, $\bm{u}=\bm{c}_\mathrm{p}^{\bm{xx}}=\bm{c}_\mathrm{p}^{\bm{yx}}=\bm{v}$, a contradiction.
Likewise, if Cases $C$ and $D$ hold, then
$\bm{c}_\mathrm{s}^{\bm{xx}}=\bm{c}_\mathrm{s}^{\bm{yy}}$, and thus $\bm{u}=\bm{c}_\mathrm{s}^{\bm{xx}}=\bm{c}_\mathrm{s}^{\bm{yx}}=\bm{v}$, also leads to a contradiction and the theorem is proved.
\end{IEEEproof}

\subsection{optimality for the quasi 2-code}
For any $\bm{x}\in \mathcal{X}^n$, we define two string operations ${\pi}_\mathcal{S}$ and $\mathrm{del}_\mathcal{S}$, where $\mathcal{S}\subset \mathbb{Z}[0,n-1]$.
Let 
$$\pi_\mathcal{S}(\bm{x})\triangleq \left\{ \bm{x}'\in \mathcal{X}^n : x'_i=x_i,\, i\in \mathbb{Z}[0,n-1]\setminus \mathcal{S} \right\}.$$ 
Let $\mathrm{del}_\mathcal{S}(\bm{x})$ be the string obtained by deleting each symbol of  $\bm{x}$ whose coordinate is in $\mathcal{S}$.
With a slight abuse of notation, let
$\mathrm{del}_\mathcal{S}(\mathcal{A}_n)=\{\mathrm{del}_\mathcal{S}(\bm{x}):\bm{x}\in \mathcal{A}_n\}$. Obviously,
$|\mathcal{A}_n|\ge |\mathrm{del}_\mathcal{S}(\mathcal{A}_n)|$.
The following lemma will be used in determining the upper bound on zero-error capacity.

\begin{lemma}\label{lm2}
For any graph $G$ representing the channel with $m$ memories (any vertex in $V(G)$ is of length $m+1$), let $\mathcal{A}_n$ be a code for $G$.
\begin{enumerate}
    \item If there exists a codeword $\bm{x}$ and a coordinate set $\mathcal{S}\subset\mathbb{Z}[0,n-1]$ such that for any
    $$j\in \mathcal{I}^m_{\mathcal{S}}\triangleq \bigcup\limits_{i\in \mathcal{S}} {\mathbb{Z}[\max\{i-m,0\},\min\{i,n-m-1\}]},$$
    the vertex $x_{j}x_{j+1}\ldots x_{j+m}$ is of degree zero, then after replacing $\bm{x}$ by any sequence in $\pi_\mathcal{S}(\bm{x})$, the updated $\mathcal{A}_n$ remains a code for $G$.
    
    \item If there exists a set $\mathcal{S}\subset\mathbb{Z}[0,n-1]$ such that for any $\bm{x},\bm{y}\in \mathcal{A}_n$ and any $j\in  \mathcal{I}^m_{\mathcal{S}}$, $x_{j}x_{j+1}\ldots x_{j+m}$ and $y_{j}y_{j+1}\ldots y_{j+m}$ are indistinguishable, then $|\mathrm{del}_\mathcal{S}(\mathcal{A}_n)|=|\mathcal{A}_n|$ and $\mathrm{del}_\mathcal{S}(\mathcal{A}_n)$ is a code for $G$.
\end{enumerate}
\end{lemma}
\begin{IEEEproof}
We first prove that when $\bm{x}$ is replaced by any sequence $\bm{x}'\in \pi_\mathcal{S}(\bm{x})$, the updated $\mathcal{A}_n$ remains a code. It is sufficient to prove that for any $\bm{y}\in \mathcal{A}_n$, $\bm{y}\neq \bm{x}$, we have $\bm{x}'$ and $\bm{y}$ are distinguishable. Since $\bm{x}$ and $\bm{y}$ are distinguishable, there exists a coordinate $j'\in\mathbb{Z}[0,n-m-1]$ such that  $x_{j'}x_{j'+1}\ldots x_{j'+m}$ and $y_{j'}y_{j'+1}\ldots y_{j'+m}$ are adjacent in $G$, which implies that the degree of $x_{j'}x_{j'+1}\ldots x_{j'+m}$ is not zero, i.e., $j'\notin \mathcal{I}_\mathcal{S}^m$.
Thus, $$\mathcal{S}\cap \mathbb{Z}[j',j'+m]=\emptyset,$$ which implies that
$x_{j'}x_{j'+1}\ldots x_{j'+m}=x'_{j'}x'_{j'+1}\ldots x'_{j'+m}$.
Therefore, $\bm{x}'$ and $\bm{y}$ are distinguishable.

Now we prove $\mathrm{del}_\mathcal{S}(\mathcal{A}_n)$ is a code for $G$. We only need to prove that for any codewords $\bm{x},\bm{y}\in \mathcal{A}_n$, $\mathrm{del}_\mathcal{S}(\bm{x})$ and $\mathrm{del}_\mathcal{S}(\bm{y})$ are distinguishable.  Since $\bm{x}$ and $\bm{y}$ are distinguishable, there exists a coordinate $j'\in\mathbb{Z}[0,n-m-1]$ such that  $x_{j'}x_{j'+1}\ldots x_{j'+m}$ and $y_{j'}y_{j'+1}\ldots y_{j'+m}$ are adjacent, which implies that $j'\notin \mathcal{I}_\mathcal{S}^m$ and so $\mathcal{S}\cap \mathbb{Z}[j',j'+m]=\emptyset$. Thus, $\mathrm{del}_\mathcal{S}(\bm{x})$ and $\mathrm{del}_\mathcal{S}(\bm{y})$ are distinguishable.
\end{IEEEproof}

\begin{figure}
\centering
\begin{tikzpicture}[scale=0.40]
    \tikzstyle{edge_style} = [line width=1.2]
    \node[above] (A) at ( 0, 0.88) {\footnotesize{$00$}};
    \node[left]  (B) at (-0.8, 0) {\footnotesize$01$};
    \node[right] (C) at ( 0.8, 0) {\footnotesize$10$};
    \node[below] (D) at ( 0,-0.88) {\footnotesize$11$};
    \fill (0,-1) circle (2pt);
    \draw[line width=1.2]
    ( 0, 1) edge ( 1, 0)
    ( 0, 1) edge ( -1, 0)
    ( -1,0) edge ( 1,0);
\end{tikzpicture}
\caption{A graph representing the channel with one memory.}\label{fig-1memory}
\end{figure}
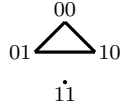

\begin{example}
For the graph $G$ in Fig.~\ref{fig-1memory}, let $\bm{x}=11001110011$ be a codeword in a code $\mathcal{A}_{11}$ for $G$. It can be seen that $\mathcal{S}=\{0,5,10\}$ and $$\mathcal{I}^m_{\mathcal{S}}=\mathbb{Z}[0,0] \cup \mathbb{Z}[4,5] \cup\mathbb{Z}[9,9]=\{0,4,5,9\}.$$
Since the degree of the vertex $11$ is zero, when the codeword $\bm{x}$ is replaced by $01001010010\in \pi_\mathcal{S}(\bm{x})$, the updated $\mathcal{A}_{11}$ remains a code for $G$. Moreover, if all the codewords in $\mathcal{A}_{11}$ start with $11$, then $\mathrm{del}_{\{0\}}(\mathcal{A}_{11})$ is also a code.
\end{example}

For disjoint sets $\mathcal{A}_i$, $i=1,2,...,$ let $\mathcal{A}_1\biguplus \mathcal{A}_2$ denote the disjoint union of $\mathcal{A}_1$ and $\mathcal{A}_2$, and $\biguplus_i \mathcal{A}_i$ denote the disjoint union of $\mathcal{A}_i$, $i=1,2,...$.

\begin{theorem}\label{thm:2-code-optimal]}
For $G(\bm{u},\bm{v})$ with $\bm{u},\bm{v}\in\mathcal{X}^{m+1}$ and $\ell(\bm{u},\bm{v})=\ell_\mathrm{p}(\bm{u},\bm{v})$, if all the symbols of $\bm{u}$ are the same, $\bm{u}=u^{m+1}$, then the capacity $C(G(\bm{u},\bm{v}))= -\log x^*$, where $x^*$ is the only positive root of
$$x^{\ell(\bm{u_v})}+x^{\ell(\bm{v_u})}=1.$$
\end{theorem}
\begin{IEEEproof}
We obtain $C(G(\bm{u},\bm{v}))\ge -\log x^*$ directly from Theorem~\ref{thm:2-code}.
Now we prove that $C(G(\bm{u},\bm{v}))\le -\log x^*$. Let $u=u_0$. By Theorem~\ref{thm:2-code}, $$\{\mathcal{B}_n\}\triangleq\{\bm{u_v},\bm{v_u}\}^*\cap \mathcal{X}^n$$ is an asymptotically optimal quasi 2-code for $G(\bm{u},\bm{v})$.
Since each prefix of $\bm{u}$ is a prefix-unit of $\bm{u}$, we have
$\ell(\bm{u_v})=m+1-\ell(\bm{u},\bm{v})$ and
  $$\bm{u_v}=u^{m+1-\ell(\bm{u},\bm{v})}.$$
On the other hand, since the first $\ell(\bm{u},\bm{v})$ symbols and the last $\ell_\mathrm{s}(\bm{u},\bm{v})$ symbols of $\bm{u}$ and $\bm{v}$ are all the same, that is, $u_i=v_i=u$ for $i\in\mathbb{Z}[0,\ell(\bm{u},\bm{v})-1]\cup \mathbb{Z}[m+1-\ell_\mathrm{s}(\bm{u},\bm{v}),m]$, we have
$$\bm{v_u}'\triangleq v_0v_1\cdots v_{m-\ell_\mathrm{s}(\bm{u},\bm{v})}$$ is a prefix-unit of $\bm{v}$. Assume that $\bm{v_u}'\neq \bm{v_u}$. We have $\ell(\bm{v_u}')> \ell(\bm{v_u})$, and then
$$u\neq v_{\ell(\bm{v_u}')-1}= v_{\ell(\bm{v_u}')-1-\ell(\bm{v_u})}.$$
Note that $v_i=u$ for $i\in\mathbb{Z}[0,\ell(\bm{u},\bm{v})-1]$. We have
$$\ell(\bm{v_u}')-1-\ell(\bm{v_u})\ge \ell(\bm{u},\bm{v})$$ and so
$$\ell(\bm{v_u})\le \ell(\bm{v_u}')-\ell(\bm{u},\bm{v})-1<m+1-\ell(\bm{u},\bm{v}),$$
which does not hold.
Therefore, $\bm{v_u}=\bm{v_u}'$ and $\ell(\bm{v_u})=m+1-\ell_\mathrm{s}(\bm{u},\bm{v})$.

Let $\{\mathcal{A}_n\}$ be a sequence of codes for $G$ such that for all $n$, $\mathcal{A}_n$ achieves the largest cardinality of a code of length $n$.
For any codeword $\bm{x}\in \mathcal{A}_n$ with $n>>m$, let
$$\mathcal{S}\triangleq \{i\in\mathbb{Z}[0,\ell(\bm{u},\bm{v})-1]\cup \mathbb{Z}[n-\ell_\mathrm{s}(\bm{u},\bm{v}),n-1]: x_i\neq u\}.$$
We can find that for any $j\in \mathcal{I}^m_{\mathcal{S}}$, the vertex $x_{j}x_{j+1}\ldots x_{j+m}$ is neither $\bm{u}$ nor $\bm{v}$, i.e., this vertex is of degree zero. By Lemma~\ref{lm2}, replacing $\bm{x}$ by $\bm{x}'\in \pi_\mathcal{S}(\bm{x})$, the updated set $\mathcal{A}_n$ remains a code, where
$$\left\{
\begin{array}{ll}
x'_i=u, &\text{if } i\in \mathcal{S}\\
x'_i=x_i, &\text{if } i\in \mathbb{Z}[0,n-1]\setminus \mathcal{S}
\end{array}\right.$$
or equivalently
$$\left\{
\begin{array}{ll}
x'_i=u, &\text{if } i\in\mathbb{Z}[0,\ell(\bm{u},\bm{v})-1]\cup \mathbb{Z}[n-\ell_\mathrm{s}(\bm{u},\bm{v}),n-1]\\
x'_i=x_i, &\text{if } i\in \mathbb{Z}[\ell(\bm{u},\bm{v}),n-1].
\end{array}\right.$$
Thus, we can assume that each codeword in $\mathcal{A}_n$ starts with $u^{\ell(\bm{u},\bm{v})}$ and ends with $u^{\ell_\mathrm{s}(\bm{u},\bm{v})}$.

Let $\bm{x}\neq u^n$ be an arbitrary but fixed sequence in the updated $\mathcal{A}_n$. Let $i$ be the first coordinate such that $x_i\neq u$. Clearly, $i\ge \ell(\bm{u},\bm{v})$.
If one of the following two conditions holds,
\begin{itemize}
\item $i+m-\ell(\bm{u},\bm{v})\ge n$ or 
\item $i+m-\ell(\bm{u},\bm{v})< n$ and $$x_ix_{i+1}\cdots x_{i+m-\ell(\bm{u},\bm{v})}\neq v_{\ell(\bm{u},\bm{v})}v_{\ell(\bm{u},\bm{v})+1}\cdots v_{m},$$
\end{itemize}
then for any
$j\in \mathcal{I}^m_{\{i\}}$, the vertex $x_{j}x_{j+1}\ldots x_{j+m}$ is neither $\bm{u}$ nor $\bm{v}$ and so of degree zero.  By Lemma~\ref{lm2}, replacing $\bm{x}$ by $\bm{x}'\in \pi_{\{i\}}(\bm{x})$, the updated set $\mathcal{A}_n$ remains a code, where
$$\left\{
\begin{array}{ll}
x'_{i'}=u, &\text{if } i'=i\\
x'_{i'}=x_{i'}, &\text{if } i'\in \mathbb{Z}[0,n-1]\setminus \{i\}.
\end{array}\right.$$
This replacement can be repeated until any codeword $\bm{x}$ in the updated $\mathcal{A}_n$ satisfies that $\bm{x}=u^n$ or
$$x_{i-\ell(\bm{u},\bm{v})}x_{i-\ell(\bm{u},\bm{v})+1}\cdots x_{i+m-\ell(\bm{u},\bm{v})}\neq \bm{v},$$
where $i$ is the first coordinate such that $x_i\neq u$.
Let
$$\mathcal{A}^i_n\triangleq \{\bm{x}\in \mathcal{A}_n: x_i\neq u \text{ and } x_{i'}=u,\forall i'\in \mathbb{Z}[0,i-1]\}$$
for $i\in\mathbb{Z}[\ell(\bm{u},\bm{v}),m]$
and $$\mathcal{A}^{m+1}_n\triangleq \{\bm{x}\in \mathcal{A}_n: x_{i'}=u,\forall i'\in \mathbb{Z}[0,m]\}.$$
Clearly,
$$\mathcal{A}_n = \biguplus_{i=\ell(\bm{u},\bm{v})}^{m+1} \mathcal{A}^i_n,$$
the disjoint union of $\mathcal{A}^i_n$, $i\in \mathbb{Z}[\ell(\bm{u},\bm{v}),m+1]$.

Note that all the sequences in $\mathcal{A}^{m+1}_n$ start with $u^{m+1}$.
Letting $\mathcal{S}_1= \mathbb{Z}[0,{m}-\ell(\bm{u},\bm{v})]$,
we can see that for any $\bm{x},\bm{y}\in \mathcal{A}^{m+1}_n$ and any $j\in  \mathcal{I}^m_{\mathcal{S}_1}=\mathcal{S}_1$, $$x_{j+\ell(\bm{u},\bm{v})}=y_{j+\ell(\bm{u},\bm{v})}=u.$$
As $v_{\ell(\bm{u},\bm{v})}\neq u$, neither $x_{j}x_{j+1}\ldots x_{j+m}$ nor $y_{j}y_{j+1}\ldots y_{j+m}$ can be $\bm{v}$, and so they are indistinguishable. Therefore, by Lemma~\ref{lm2}, we have
\begin{equation}\label{eq:2-code-optimal-1}
|\mathcal{A}^{m+1}_n|=|\mathrm{del}_{\mathcal{S}_1}(\mathcal{A}^{m+1}_n)|\le |\mathcal{A}_{n-|\mathcal{S}_1|}|=|\mathcal{A}_{n-(m+1-\ell(\bm{u},\bm{v}))}|.
\end{equation}

Now we consider $\mathcal{A}'_n\triangleq \bigcup_{i=\ell(\bm{u},\bm{v})}^{m} \mathcal{A}^i_n$.
For any sequence $\bm{a}\in \mathcal{A}'_n$, there exists an $i\in\mathbb{Z}[\ell(\bm{u},\bm{v}),m]$ such that $\bm{a}\in \mathcal{A}^i_n$.
Then $a_i\neq u$. Letting $\mathcal{S}_2= \mathbb{Z}[0,{m}-\ell_\mathrm{s}(\bm{u},\bm{v})]$, we can see that for any $j\in  \mathcal{I}^m_{\mathcal{S}_2}=\mathcal{S}_2$,
$$a_{j}a_{j+1}\ldots a_{j+m}\neq \bm{u}.$$
Therefore,  for any $\bm{x},\bm{y}\in \mathcal{A}^{m+1}_n$ and any $j\in  \mathcal{I}^m_{\mathcal{S}_2}$, neither $x_{j}x_{j+1}\ldots x_{j+m}$ nor $y_{j}y_{j+1}\ldots y_{j+m}$ can be $\bm{u}$, and so they are indistinguishable.
By Lemma~\ref{lm2}, we have
\begin{equation}\label{eq:2-code-optimal-2}
|\mathcal{A}'_n|=|\mathrm{del}_{\mathcal{S}_2}(\mathcal{A}'_n)|\le |\mathcal{A}_{n-|\mathcal{S}_2|}|=|\mathcal{A}_{n-(m+1-\ell_\mathrm{s}(\bm{u},\bm{v}))}|.
\end{equation}
By (\ref{eq:2-code-optimal-1}) and (\ref{eq:2-code-optimal-2}), we have
\begin{equation*}
\begin{split}		
    |\mathcal{A}_n|
    &\le |\mathcal{A}_{n-(m+1-\ell(\bm{u},\bm{v}))}|+|\mathcal{A}_{n-(m+1-\ell_\mathrm{s}(\bm{u},\bm{v}))}|\\
    &=|\mathcal{A}_{n-\ell(\bm{u_v})}|+|\mathcal{A}_{n-\ell(\bm{v_u})}|,
\end{split}
\end{equation*}
which is a classic recurrence formula. We have
$C(G)=R(\{\mathcal{A}_n\})\le -\log x^*$.
\end{IEEEproof}

\section{Capacity of the Binary Channel with Two Memories}\label{sec:twomemories}
In this section, we consider all the 28 graphs with one edge, which can be classified in 11 cases and have been listed in Table~\ref{table}. As discussed in Section~I, for each case, we only need to consider any one of the graphs therein.
The zero-error capacity of the graphs in Cases~1 to 10 are solved in this paper.
We will also give a lower bound and a upper bound on the zero-error capacity of the graphs in Case~11.

According to Theorem~\ref{thm:2-code-optimal]}, Theorems~\ref{thm:1} to \ref{thm:3} can be obtained immediately. The proofs are omitted.
\begin{theorem}\label{thm:1}
$C(G)=-\log \alpha \approx 0.551$ for $G=G(000,001)$, where $\alpha$ is the only positive root of the equation
$$x  +x^3=1.$$
\end{theorem}
\begin{theorem}\label{thm:2}
$C(G)=\frac{1}{2}$ for $G=G(000,010)$.
\end{theorem}
\begin{theorem}\label{thm:3}
$C(G)=-\log \beta \approx 0.406$ for $G=G(000,011)$, where $\beta$ is the only positive root of the equation
$$x^2+x^3=1.$$
\end{theorem}

To facilitate the proofs of the following theorems, for a set of sequences $\mathcal{A}_n$ of length $n$, and a string $\bm{x}$ with length strictly less than $n$, we denote $\mathcal{A}_n^{\bm{x}}$ be the subset of sequences in $\mathcal{A}_n$ starting with $\bm{x}$.

\begin{theorem}\label{thm:4}
$C(G)=-\log \beta \approx 0.406$ for $G=G(010,011)$, where $\beta$ is the only positive root of the equation
$$x^2+x^3=1.$$
\end{theorem}
\begin{IEEEproof}
By Theorem~\ref{thm:2-code}, we obtain that $C(G)\ge -\log \beta.$

To prove $C(G)\le -\log \beta$,
let $\{\mathcal{A}_n\}$ be a sequence of codes for $G$ such that for all $n$, $\mathcal{A}_n$ achieves the largest cardinality of a code of length $n$.
Flipping the first bit of any codeword in $\mathcal{A}_n$ starting with $1$, and the second bit of any codeword starting with $000$, by Lemma~\ref{lm2}, the updated $\mathcal{A}_n$ remains a code. Thus, WLOG, we assume that any codeword in $\mathcal{A}_n$ starts with $011$, $010$ or $001$. Equivalently, we assume that any codeword in $\mathcal{A}_n$ starts with $011$, $010$, $0010$ or $0011$, i.e., $$\mathcal{A}_n=\mathcal{A}_n^{011}\biguplus \mathcal{A}_n^{010}\biguplus \mathcal{A}_n^{0010}\biguplus \mathcal{A}_n^{0011},$$
and so
\begin{equation}\label{eqb1}
    |\mathcal{A}_n|=\left|\mathcal{A}_n^{011}\biguplus \mathcal{A}_n^{0011}\right|+\left|\mathcal{A}_n^{010}\biguplus \mathcal{A}_n^{0010}\right|.
\end{equation}	
Obviously,
\begin{equation}\label{eqb2}
    \begin{split}
        \left|\mathcal{A}_n^{011}\biguplus \mathcal{A}_n^{0011}\right|&=\left|\mathrm{del}_{\{0,1,2\}}\left(\mathcal{A}_n^{011}\biguplus \mathcal{A}_n^{0011}\right)\right|,\\
        \left|\mathcal{A}_n^{010}\biguplus \mathcal{A}_n^{0010}\right|&=\left|\mathrm{del}_{\{0,1\}}\left(\mathcal{A}_n^{010}\biguplus \mathcal{A}_n^{0010}\right)\right|,
    \end{split}
\end{equation}
and by Lemma~\ref{lm2}, both $\mathrm{del}_{\{0,1,2\}}\left(\mathcal{A}_n^{011}\biguplus \mathcal{A}_n^{0011}\right)$ and $\mathrm{del}_{\{0,1\}}\left(\mathcal{A}_n^{010}\biguplus \mathcal{A}_n^{0010}\right)$ are codes for $G$. 
Note that both $\mathcal{A}_{n-3}$ and $\mathcal{A}_{n-2}$ achieve maximum cardinality for codes of lengths $n-3$ and $n-2$, respectively. We have
\begin{equation}\label{eqb3}
    \begin{split}
        \left|\mathrm{del}_{\{0,1,2\}}\left(\mathcal{A}_n^{011}\biguplus \mathcal{A}_n^{0011}\right)\right|&\le |\mathcal{A}_{n-3}|,\\
        \left|\mathrm{del}_{\{0,1\}}\left(\mathcal{A}_n^{010}\biguplus \mathcal{A}_n^{0010}\right)\right|&\le |\mathcal{A}_{n-2}|.
    \end{split}
\end{equation}
By \eqref{eqb1}-\eqref{eqb3}, we have
$|\mathcal{A}_n|\le |\mathcal{A}_{n-2}| + |\mathcal{A}_{n-3}|$,
which is a classic recurrence formula. Therefore,
$C(G)=R(\{\mathcal{A}_n\})\le -\log \beta$.
\end{IEEEproof}

\begin{theorem}\label{thm:5}
$C(G)=-\log \beta \approx 0.406$ for $G=G(010,001)$, where $\beta$ is the only positive root of the equation
$$x^2+x^3=1.$$
\end{theorem}
\begin{IEEEproof}
By Theorem~\ref{thm:2-code}, we obtain that $C(G)\ge -\log \beta.$

To prove $C(G)\le -\log \beta$,
let $\{\mathcal{A}_n\}$ be a sequence of codes for $G$ such that for all $n$, $\mathcal{A}_n$ achieves the largest cardinality of a code of length $n$.
Flip the first bit of any codeword in $\mathcal{A}_n$ starting with $1$. For any codeword containing $11$, replace any one of 11s by 10. By Lemma~\ref{lm2}, the updated $\mathcal{A}_n$ remains a code. Thus, we assume that any codeword in $\mathcal{A}_n$ starts with $001$, $000$ or $010$, i.e., $$\mathcal{A}_n=\mathcal{A}_n^{001}\biguplus \mathcal{A}_n^{000}\biguplus \mathcal{A}_n^{010},$$
and so
\begin{equation*}\label{eqc1}
|\mathcal{A}_n|=\left|\mathcal{A}_n^{001}\right|+\left|\mathcal{A}_n^{000}\biguplus \mathcal{A}_n^{010}\right|.
\end{equation*}	
We also have
\begin{equation*}\label{eqc2}
    \begin{split}
        \left|\mathcal{A}_n^{001}\right|&=|\mathrm{del}_{\{0,1,2\}}(\mathcal{A}_n^{001})|\\
        \left|\mathcal{A}_n^{000}\biguplus \mathcal{A}_n^{010}\right|&=\left|\mathrm{del}_{\{0,1\}}(\mathcal{A}_n^{000}\biguplus \mathcal{A}_n^{010})\right|,
    \end{split}
\end{equation*}
and by Lemma~\ref{lm2}, both $\mathrm{del}_{\{0,1,2\}}(\mathcal{A}_n^{001})$ and $\mathrm{del}_{\{0,1\}}(\mathcal{A}_n^{000}\biguplus \mathcal{A}_n^{010})$ are codes for $G$. Thus,
$|\mathcal{A}_n|\le |\mathcal{A}_{n-2}| + |\mathcal{A}_{n-3}|$, which implies that $C(G)=R(\{\mathcal{A}_n\})\le -\log \beta$.
\end{IEEEproof}

\begin{lemma}\label{thm:0}
$C(G)=\frac{1}{3}$ for $G$ containing all four edges $\{000,111\}$, $\{010,101\}$, $\{100,011\}$ and $\{110,001\}$.
\end{lemma}
\begin{IEEEproof}
We can easily obtain a sequence of codes for $G$:
$$\mathcal{A}_n^*\triangleq\{000,111\}^*\bigcap \{0,1\}^n.$$
whose rate is $R(\{\mathcal{A}_n^*\})=\frac{1}{3}.$

Now we consider the proof of the upper bound.
Let $\{\mathcal{A}_n\}$ be a sequence of codes for $G$ such that for all $n$, $\mathcal{A}_n$ achieves the largest cardinality of a code of length $n$.
Obviously,
\begin{equation*}
    \begin{split}
        |\mathcal{A}_n|&=\sum_{i_0,i_1,i_2\in\{0,1\}}|\mathcal{A}_n^{i_0i_1i_2}|\\
             &=\sum_{i_2\in\{0,1\}}\left|\biguplus_{i_0,i_1\in\{0,1\}} \mathcal{A}_n^{i_0i_1i_2}\right|.
    \end{split}
\end{equation*}	
As for any $i_2\in \{0,1\}$, the third bits of any two codewords in $\biguplus_{i_0,i_1\in\{0,1\}} \mathcal{A}_n^{i_0i_1i_2}$ are the same, by Lemma~\ref{lm2}, we see that $\mathrm{del}_{\{0,1,2\}}\left(\biguplus_{i_0,i_1\in\{0,1\}} \mathcal{A}_n^{i_0i_1i_2}\right)$ is also a code for $G$ and
\begin{equation*}
\begin{split}
\left|\biguplus_{i_0,i_1\in\{0,1\}} \mathcal{A}_n^{i_0i_1i_2}\right|&=\left|\mathrm{del}_{\{0,1,2\}}\left(\biguplus_{i_0,i_1\in\{0,1\}} \mathcal{A}_n^{i_0i_1i_2}\right)\right|\\
&\le |\mathcal{A}_{n-3}|.
\end{split}
\end{equation*}	
Thus,
$|\mathcal{A}_n|\le 2|\mathcal{A}_{n-3}|$,
i.e., $C(G)=R(\{\mathcal{A}_n\})\le 1/3$.
\end{IEEEproof}

The following lemma is evident, and so the proof is omitted.
\begin{lemma}\label{lm:1/m+1}
For $G(\bm{u},\bm{v})$ with $\bm{u},\bm{v}\in\mathcal{X}^{m+1}$, $\{\bm{u},\bm{v}\}^{k}$ is a code for each $k$, and $C(G)\ge \frac{1}{m+1}$.
\end{lemma}

According to Lemmas~\ref{thm:0} and \ref{lm:1/m+1}, we can obtain Theorems~\ref{thm:6} to \ref{thm:8} immediately.
\begin{theorem}\label{thm:6}
$C(G)=\frac{1}{3}$ for $G=G(000,111)$.
\end{theorem}

\begin{theorem}\label{thm:7}
$C(G)=\frac{1}{3}$ for $G=G(010,101)$.
\end{theorem}

\begin{theorem}\label{thm:8}
$C(G)=\frac{1}{3}$ for $G=G(100,011)$.
\end{theorem}

\begin{theorem}\label{thm:9}
$C(G)=\frac{1}{3}$ for $G=G(000,101)$.
\end{theorem}
\begin{IEEEproof}
We can easily obtain a sequence of codes for $G$:
$$\mathcal{A}_n^*\triangleq\{000,101\}^*\bigcap \{0,1\}^n$$
whose rate is $R(\{\mathcal{A}_n^*\})=\frac{1}{3}.$

Let $G'$ be the graph with two edges $\{000,101\}$ and $\{000,111\}$. Since $E(G)\subseteq E(G')$, we have $C(G)\le C(G')$. Let $\{\mathcal{A}_n\}$ be asymptotically optimal for $G'$.

If no codeword in $\mathcal{A}_n$ contains the substrings $101$, then $\mathcal{A}_n$ is a code for $G(000,111)$. Otherwise, we perform a sequence of substring replacements. Specifically, let $\bm{x}\in \mathcal{A}'_n$ be an arbitrary codeword which contains the substring 111. Then replace any one of 111s by 101. The updated $\mathcal{A}_n$ remains a code. Thus the replacement
can be repeated until no codeword in the updated $\mathcal{A}_n$ contains the substring 101, and therefore the final updated $\mathcal{A}_n$ is a code for $G(000,111)$.
Thus, $C(G)\le C(G')=R(\{\mathcal{A}_n\})\le C(G(000,111))=\frac{1}{3}.$
\end{IEEEproof}

\begin{theorem}\label{thm:10}
$C(G)=\frac{1}{3}$ for $G=G(001,011)$.
\end{theorem}
\begin{IEEEproof}
By Theorem~\ref{thm:2-code}, we obtain that $C(G)\ge -\log \beta.$

The proof of the upper bound is also similar to the proof of Theorem~\ref{thm:6}.
Let $\{\mathcal{A}_n\}$ be a sequence of codes for $G$ such that for all $n$, $\mathcal{A}_n$ achieves the largest cardinality of a code of length $n$.
Flip the first bit of any codeword in $\mathcal{A}_n$ starting with $1$ and the second bit of any codeword starting with $0101$. For any codeword containing $111$ or $0000$, replace any one of 111s by 110 or 0000s by 0100. By Lemma~\ref{lm2}(1), the final updated $\mathcal{A}_n$ remains a code. Thus, we can assume that any codeword in $\mathcal{A}_n$ starts with 0001, 0010, 0011, 0100 or 0110, i.e., $$\mathcal{A}_n=\mathcal{A}_n^{0001}\biguplus \mathcal{A}_n^{0010}\biguplus \mathcal{A}_n^{0011}\biguplus \mathcal{A}_n^{0100}\biguplus \mathcal{A}_n^{0110},$$
and so
\begin{equation*}\label{eqd1}
    |\mathcal{A}_n|=\left|\mathcal{A}_n^{0100}\biguplus \mathcal{A}_n^{0010}\biguplus \mathcal{A}_n^{0011}\right|+\left|\mathcal{A}_n^{0110}\biguplus \mathcal{A}_n^{0001}\right|.
\end{equation*}	
We also have
\begin{equation*}\label{eqd2}
    \begin{split}
        &\left|\mathcal{A}_n^{0100}\biguplus \mathcal{A}_n^{0010}\biguplus \mathcal{A}_n^{0011}\right|=\\
        &\left|\mathrm{del}_{\{0,1,2\}}(\mathcal{A}_n^{0100}\biguplus \mathcal{A}_n^{0010}\biguplus \mathcal{A}_n^{0011})\right|
    \end{split}
\end{equation*}
and
\begin{equation*}
    \left|\mathcal{A}_n^{0110}\biguplus \mathcal{A}_n^{0001}\right|=\left|\mathrm{del}_{\{0,1,2\}}(\mathcal{A}_n^{0110}\biguplus \mathcal{A}_n^{0001})\right|.
\end{equation*}
Both $\mathrm{del}_{\{0,1,2\}}(\mathcal{A}_n^{0100}\biguplus \mathcal{A}_n^{0010}\biguplus \mathcal{A}_n^{0011})$ and $\mathrm{del}_{\{0,1,2\}}(\mathcal{A}_n^{0110}\biguplus \mathcal{A}_n^{0001})$ are codes for $G$. Thus,
$|\mathcal{A}_n|\le 2|\mathcal{A}_{n-3}|,$
i.e.,
$C(G)=R(\{\mathcal{A}_n\})\le \frac{1}{3}$.
\end{IEEEproof}

\begin{theorem}\label{thm:11}
$\frac{\log 14}{11}\le C(G)\le -\log \beta\approx 0.406$ for $G=G(001,100)$, where $\beta$ is the only positive root of the equation
$$x^2+x^3=1.$$
\end{theorem}
\begin{IEEEproof}
We can obtain a sequence of codes for $G$:
\begin{multline*}
    \mathcal{A}_n^*=\{00100100100,00100101001,00100110010,
00110010010,00110011001,01001001001,01001001100,\\
01001100100,10000100001,10010010010,10010011001,
10010100100,10011001001,10011001100\}^*\bigcap \{0,1\}^n,
\end{multline*}
whose rate is $R(\{\mathcal{A}_n^*\})=\frac{\log 14}{11}.$

The proof of the upper bound is also similar to the proof of Theorem~\ref{thm:2}.
Let $\{\mathcal{A}_n\}$ be a sequence of codes for $G$ such that for all $n$, $\mathcal{A}_n$ achieves the largest cardinality of a code of length $n$.
Flip the first bit of any codeword in $\mathcal{A}_n$ starting with $000$, $110$ or $101$, the second bit of any codeword starting with $011$, and the first two bits of any codeword starting with $111$. Thus, we can assume that any codeword in $\mathcal{A}_n$ starts with 001, 010, 100, i.e., $$\mathcal{A}_n=\mathcal{A}_n^{001}\biguplus \mathcal{A}_n^{010}\biguplus \mathcal{A}_n^{100},$$
and so
\begin{equation*}\label{eqd1}
    |\mathcal{A}_n|=|\mathcal{A}_n^{100}|+\left|\mathcal{A}_n^{001}\biguplus \mathcal{A}_n^{010}\right|.
\end{equation*}	
By Lemma~\ref{lm2}(2), we have
    \vspace{-1mm}
$$|\mathcal{A}_n^{100}|=|\mathrm{del}_{\{0,1,2\}}(\mathcal{A}_n^{100})|$$
and
$$\left|\mathcal{A}_n^{001}\biguplus \mathcal{A}_n^{010}\right|=\left|\mathrm{del}_{\{0,1\}}(\mathcal{A}_n^{001}\biguplus \mathcal{A}_n^{010})\right|.$$
Both $\mathrm{del}_{\{0,1,2\}}(\mathcal{A}_n^{100})$ and $\mathrm{del}_{\{0,1\}}(\mathcal{A}_n^{001}\biguplus \mathcal{A}_n^{010})$ are codes for $G$. Thus,
$$|\mathcal{A}_n|\le |\mathcal{A}_{n-3}|+|\mathcal{A}_{n-2}|,$$
i.e.,
$C(G)=R(\{\mathcal{A}_n\})\le -\log \beta$.
\end{IEEEproof}

\begin{remark}
We can see that for the graph in Cases 1 to 10, the optimal quasi 2-code constructed in Theorem~\ref{thm:2-code} achieves the zero-error capacity.
However, for $G=G(001,100)$ (Case 11), the optimal quasi 2-code is $\{001,100\}^*\bigcap \{0,1\}^n$ whose rate is $\frac{1}{3}$.
The capacity $C(G)\ge \frac{\log 14}{11}>\frac{1}{3}$.
\end{remark}

\section{Conclusion}
In this paper, we have investigated the zero-error capacity of channels characterized by graphs containing a single edge.
Previous works primarily focused on binary input channels with 2 or 3 memories. Our study extends the analysis to channels with $|\mathcal{X}|$-ary inputs and arbitrary memories.
We provide a method for constructing zero-error codes for such graphs with one edge, which offers a lower bound on the zero-error capacity. For the binary channel two memories, the zero-error codes constructed by this method have been proven to be optimal in most cases.

It could be valuable to construct a method to obtain a general upper bound for graphs with one edge. If this upper bound matches the lower bound derived in this paper, it would allow us to determine the capacity for numerous graphs of this type.

\bibliographystyle{IEEEtran}
\bibliography{References}

\end{document}